\documentclass[linenumbers]{aa}

\usepackage{graphicx}
\usepackage{txfonts}
\usepackage[dvipsnames]{xcolor}
\usepackage{tabularx}
\usepackage{chemformula}
\usepackage{physics}
\usepackage{scrextend}
\usepackage{makecell}
\usepackage[pdfpagelabels=false]{hyperref}	
\hypersetup{colorlinks=true,linkcolor=blue,citecolor=blue,filecolor=blue,urlcolor=blue,}
\usepackage{placeins}

\def\figurenum#1{\def\thefigure{#1}\let\currentlabel\thefigure}

\usepackage{siunitx}
\usepackage{threeparttable} 

\begin{document} 

\title{The Retrieved Atmospheric Properties of the Sub-stellar Object VHS~1256~b from Analyzing HST, VLT and JWST Spectra}
\titlerunning{Atmospheric Retrieval Applied to Variable VHS~1256~b}
\authorrunning{Lueber et al.}

\author{Anna Lueber\inst{1,2}$^{\href{}{}}$
         \and Kevin Heng\inst{1,3,4,5}
         \and  Brendan P. Bowler\inst{6}
         \and  Daniel Kitzmann\inst{2,7}
         \and  Johanna M. Vos\inst{8}
         \and  Yifan Zhou\inst{9}}
\institute{Ludwig Maximilian University, Faculty of Physics, University Observatory, Scheinerstr. 1, Munich D-81679, Germany\\
\email{anna.lueber@physik.lmu.de, kevin.heng@physik.lmu.de}
\and University of Bern, Center for Space and Habitability, Gesellschaftsstrasse 6, CH-3012, Bern, Switzerland
\and ARTORG Center for Biomedical Engineering Research, University of Bern, Murtenstrasse 50, CH-3008, Bern, Switzerland
\and University College London, Department of Physics \& Astronomy, Gower St, London, WC1E 6BT, United Kingdom
\and University of Warwick, Department of Physics, Astronomy \& Astrophysics Group, Coventry CV4 7AL, United Kingdom
\and The University of Texas at Austin, Department of Astronomy, 2515 Speedway, Stop C1400, Austin, TX 78712, U.S.A.
\and Space Research and Planetary Sciences, Physics Institute, University of Bern, Gesellschaftsstrasse 6, 3012 Bern, Switzerland
\and School of Physics, Trinity College Dublin, The University of Dublin, College Green, Dublin 2, Ireland
\and University of Virginia, Department of Astronomy, 530 McCormick Rd, Charlottesville, VA 22904, U.S.A.
}

\date{Received ; accepted}

\abstract{Motivated by the observed $\sim 30\%$ variations in flux from the L7 dwarf VHS~1256~b, we subjected its time-resolved \textit{Hubble} Space Telescope (HST) WFC3 spectra (measured in two epochs in 2018 and 2020), as well as medium-resolution Very Large Telescope (VLT) X-shooter and Early Release Science \textit{James Webb} Space Telescope (JWST) spectra to a suite of both standard Bayesian (nested sampling) and machine-learning (random forest) retrievals. We find that both HST and VLT data require vertically varying abundance profiles of water in order to model the spectra accurately. Despite the large flux variations observed in the HST data, the temporal variability cannot be attributed to a single varying atmospheric property.  The retrieved atmospheric quantities are consistent with being invariant across time.  However, we find that model grids provide generally poor fits to the measured HST spectra and are unsuitable for quantifying the temporal variability of atmospheric properties.  Additionally, our analysis of JWST spectra using model grids indicates consistency in retrieved properties across different wavelength channels.  Despite the temporal variability in flux, the retrieved properties between HST and VLT, as well as between HST and JWST, are consistent within the respective posterior uncertainties. Such an outcome bodes well for future retrieval analyses of exoplanetary atmospheres, which are expected to exhibit weaker flux variations.}


\maketitle
\section{Introduction}
\label{sect:intro}

The study of variability in the atmospheres of planets and sub-stellar objects has a long and rich history, e.g., the observed variability of the atmosphere of Earth over a wide variety of timescales \citep{Peixoto1984RvMP...56..365P}.  The variability of brown dwarfs at the $\sim 1\%$ level has been observationally established for over a decade \citep{Artigau2009ApJ...701.1534A, Apai2013ApJ, Radigan2014ApJ, Metchev2015ApJ...799..154M}.  It is believed to be caused by rotational modulated surface features that are the consequence of heterogeneous cloud cover \citep{Showman2013ApJ...776...85S, Apai2013ApJ, Crossfield2014Natur.505..654C, Morley2014ApJ}.  Several time series observations support this idea (e.g., \citealt{Radigan2012ApJ...750..105R, Buenzli2014ApJ...782...77B, Metchev2015ApJ...799..154M}).  More recent studies have suggested that these clouds possess vertical structures \citep{Apai2013ApJ, Yang2015ApJ...798L..13Y, Zhou2018AJ....155..132Z}. Heterogeneities in thermal structures \citep{Robinson2014ApJ...785..158R} and methane abundance \citep{Tremblin2020A&A...643A..23T} have also been proposed. Additionally, a recent study by \cite{Rowland2023ApJ...947....6R} examined the impact of abundance profile assumptions by performing atmospheric retrievals on synthetic, cloud-free L dwarf spectra generated from the Sonora Bobcat models at SpeX resolution. They demonstrated that a non-uniform (step function) description for the iron hydride abundance profiles is needed to accurately retrieve bulk properties of early L dwarfs.

The study of brown dwarf variability is an important precursor to the study of variability in exoplanetary atmospheres, which is still in its infancy and reports of variability remain debated (e.g., \citealt{Agol2010ApJ...721.1861A, Apai2016ApJ...820...40A, Armstrong2016NatAs...1E...4A, Cowan2017MNRAS.467..747C, Jackson2019AJ....157..239J, Hooton2019MNRAS.486.2397H,Komacek2020ApJ...888....2K, Biller2021, Cho2021ApJ...913L..32C, Jones2022A&A...666A.118J, Hardy2022ApJ...940..123H, Lally2022AJ....163..181L}).

VHS~1256~b (full name: VHS J125060.192-125723.9 b), originally discovered by \citet{Gauza2015ApJ...804...96G}, has a spectral type of L7, a distance of 21.14 $\pm$ 0.22 pc \citep{Gaia2016AA, Gaia2021A&A...649A...1G}, an age of $140 \pm 20$ Myr \citep{Dupuy2023MNRAS.519.1688D}, and a low surface gravity \citep{Petrus2023A&A...670L...9P}. Recently, \cite{Zhou2022AJ....164..239Z} analyzed Wide Field Camera 3 (WFC3) spectra of VHS~1256~b from 1.12 to 1.65 $\mu$m using the \textit{Hubble} Space Telescope (HST) at two different epochs \citep{Bowler2020ApJ...893L..30B, Zhou2022AJ....164..239Z}. The object stands out as a prime example of a variable brown dwarf as it exhibits 24\% color variability on the timescale of a rotation period ($\sim$22 hr). Additionally, for long-term variability on a timescale of nearly $\sim$900 rotation periods, a peak-to-valley flux difference of 33\%$\,\pm \,$2\% was found, with an even higher amplitude reaching 38\% in the J band, the highest amplitude ever observed in a sub-stellar object \citep{Bowler2020ApJ...893L..30B, Zhou2022AJ....164..239Z}. 

The luminosity of VHS~1256~b at its current age is consistent with two possible evolutionary tracks: one where deuterium is inert, and another where it fuses. As a result, there are two potential mass estimates for VHS~1256~b, 12.0$\,\pm\,$0.1~M$_\mathrm{Jup}$ or 16$\,\pm\,$1~M$_\mathrm{Jup}$, based on evolutionary models from \cite{SaumonMarley2008ApJ}, which produces a bimodal probability distribution that places it on either side of the exoplanet-brown dwarf division. The corresponding surface gravities $\log g$, with $g$ in cgs units, were recently constrained by \cite{Dupuy2023MNRAS.519.1688D}. They reported values of $\log{g}=4.268 \pm 0.006$ and $\log{g}=4.45 \pm 0.03$, respectively.

Near-IR spectra of VHS~1256~b have revealed weak alkali absorption, suggestive of its youth \citep{Gauza2015ApJ...804...96G, Petrus2023A&A...670L...9P, Dupuy2023MNRAS.519.1688D}. Fitting atmospheric forward models \citep{Burrows2006ApJ...640.1063B} to spectrophotometric data suggests the presence of a cloudy atmosphere in VHS~1256~b. The observed variability suggests dynamic atmospheric processes, potentially influenced by a patchy cloud structure, as demonstrated by \citet{Vos2023ApJ...944..138V} for two variable, planetary-mass early T dwarfs using HST and Spitzer data. Furthermore, weak methane absorption suggests a relative scarcity of methane, which may be attributed to disequilibrium chemistry \citep{Miles2023ApJ...946L...6M}. Large variability of NIR-spectra on the timescale of a rotation period motivates the question of whether the atmospheric properties of VHS~1256~b are changing on these timescales---and if so, which are the most variable properties.

The traditional approach of analyzing brown dwarf spectra is to extract its atmospheric and bulk properties. This process involves performing interpolation on a pre-computed model grid of spectra and evolutionary tracks, and subsequently acquiring the best-fit spectra through the application of a goodness-of-fit metric (e.g., \citealt{Marley1996Sci, Burrows1997ApJ...491..856B, Chabrier2000ApJ, AckermanMarley2001ApJ...556..872A, Allard2001ApJ...556..357A, Baraffe2002A&A, Burrows2003ApJ, Burrows2011ApJ, Morley2014ApJ, Zhang2021ApJ, Zhang2021bApJ...921...95Z}).  This allows the practitioner to infer the surface gravity, mass and effective temperature of the brown dwarf (e.g., \citealt{Burgasser2006ApJ...639.1095B, Cushing2008ApJ...678.1372C, Kasper2009ApJ...695..788K, King2010, Bowler2010ApJ...723..850B, Hinkley2015ApJ...805L..10H, Franson2023AJ....165...39F}). The caveat to such an approach is that the model parameters are usually sampled uniformly (either in a linear or logarithmic sense), which may lead to biased posterior distributions when sampling over more than two parameters, particularly for parameters exhibiting nonlinear effects on the spectrum \citep{Fisher2022ApJ...934...31F}. It has also been shown that these self-consistent model grids, despite incorporating an array of sophisticated chemistry and physics, systematically under- or overestimate the inferred mass of brown dwarfs for objects where dynamical masses have been measured \citep{Dupuy2009ApJ...692..729D, Konopacky2010ApJ...711.1087K, Cheetham2018A&A...614A..16C, Beatty2018AJ....156..168B, Rickman2020A&A...635A.203R, Bowler2021ApJ...913L..26B, Brandt2021ApJ...915L..16B}.  

In more recent years, techniques have been developed applying machine learning methods to analyze the atmospheres of exoplanets and brown dwarfs (e.g., \citealt{Waldmann2016ApJ...820..107W, Marquez2018NatAs...2..719M, Zingales2018AJ....156..268Z, Cobb2019AJ....158...33C, Fisher2020AJ....159..192F, Yip2021AJ....162..195Y, Matchev2022ApJ...930...33M, Ardevol2022A&A...662A.108A, Vasist2023A&A...672A.147V, Gebhard2023arXiv231208295G}). Instead of interpolating on a pre-computed grid of models, machine learning methods use the model grid as a training set, which is used to map a measured spectrum to the parameter values of the grid.  Traditional Bayesian methods often sacrifice physical realism for computational feasibility and are typically not self-consistent, but are capable of exploring parameter space widely. Therefore, while we will utilize self-consistent model grids in our interpretation of the HST WFC3, VLT/X-shooter, and ERS JWST spectra of VHS~1256~b, in tandem with a supervised machine learning approach \citep{Marquez2018NatAs...2..719M}, we will also perform the interpretation of spectra using the standard Bayesian inference approach of nested sampling \citep{Skilling2006, Feroz2009MNRAS, Trotta2008ConPh, Kitzmann2020ApJ} for HST WFC3 and VLT/X-shooter. 

In Section~\ref{sect:Methodology}, we describe the archival data set that we are using for this study, the atmospheric retrieval techniques as well as the legacy model grids used for the current study. Outcomes from a large number of atmospheric retrievals, both orbit-averaged and intra-orbit, as well as their time series analysis for HST data are presented in Section~\ref{sect:Results}. This section also includes the \texttt{BeAR} retrieval analysis of the curated VLT/X-shooter spectrum and grid-based machine-learning retrievals for the ERS JWST spectrum using \texttt{HELA}. Section~\ref{sect:Discussion} presents implications from our study and opportunities for future work.

\section{Methodology}
\label{sect:Methodology}

\subsection{Data}
\label{sect:Data}

\begin{figure}
    \centering
    \includegraphics[width=\linewidth]{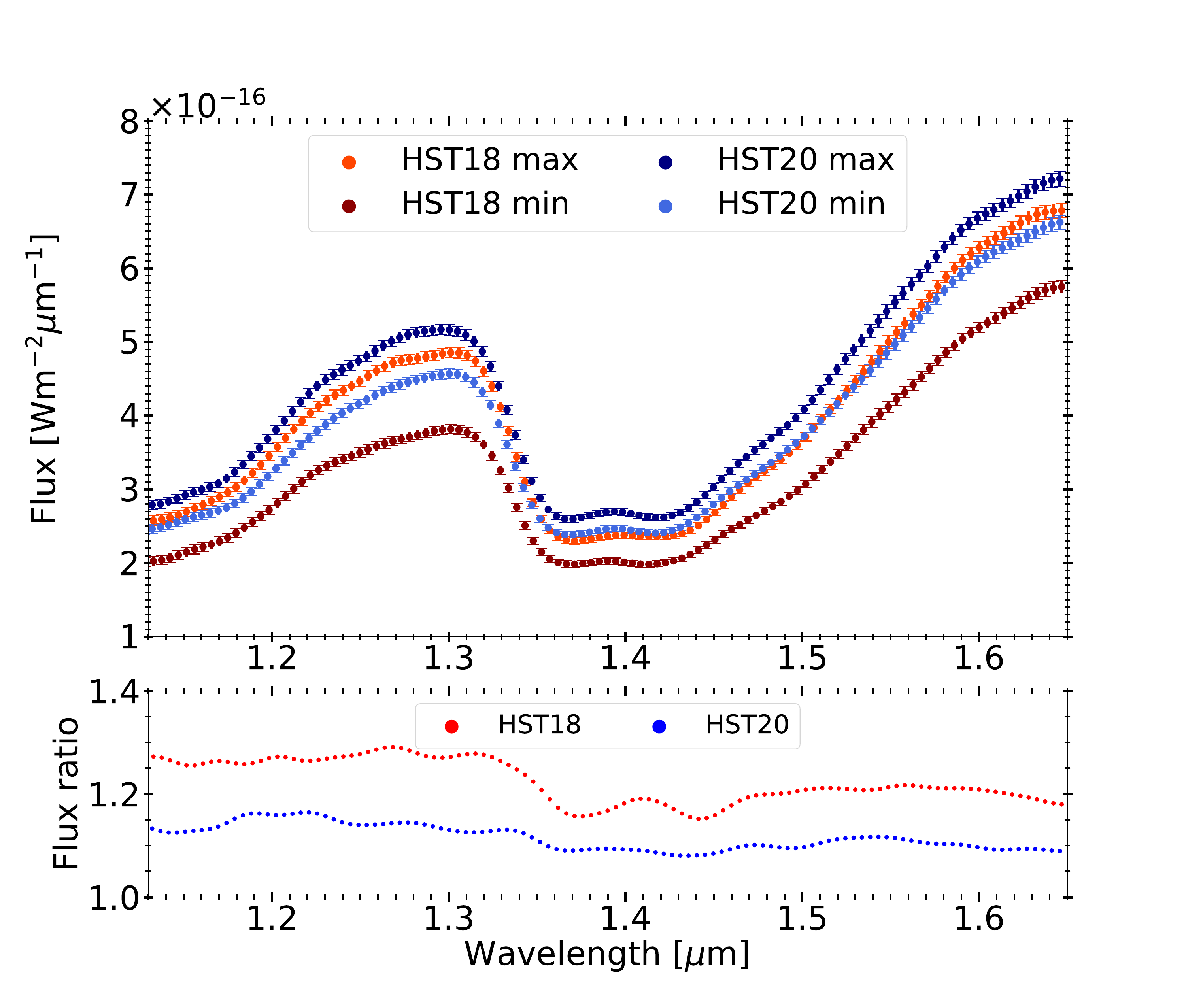}
    \caption{Minimum and maximum fluxes from the two epochs of HST observations of VHS~1256~b in 2018 and 2020.  The data was first published by \cite{Bowler2020ApJ...893L..30B} and \cite{Zhou2022AJ....164..239Z}, and are displayed here for illustration.}
    \label{fig:HST18vsHST20}
\end{figure}

The two epochs of HST-WFC3 spectra of VHS~1256~b, taken in 2018 and 2020, were originally published in \cite{Bowler2020ApJ...893L..30B} and \cite{Zhou2022AJ....164..239Z}.  The six-orbit continuous monitoring from UT 2018-03-05 16:02:30 to 2018-03-06 00:42:47 (Program ID: GO-15197; PI: Bowler) resulted in a light curve that spanned less than half of its rotation period ($\sim 9$\,hr). In 2020, spectroscopic monitoring of VHS~1256~b was performed for fifteen orbits from UT 2020-05-26 09:27:17 to 2020-05-28 03:29:24 (Program ID: GO- 16036; PI: Zhou), covering approximately two rotation periods ($\sim 42$\,hr).  The spectral resolution is R$\sim$130 at 1.4 $\mu$m and the wavelength range is from $1.12\,\mu$m to $1.65\,\mu$m.

\begin{table*}[ht!]
\centering
\caption{Summary of retrieval parameters and prior distributions for the free chemistry approach used in the cloud-free, gray cloud and non-gray cloud models.}
\label{tab:priors}
\resizebox{0.75\textwidth}{!}{
\begin{threeparttable}
\begin{tabular}{lcccc}
\hline
Parameter & Symbol & Prior Type & & Prior Value\\
\hline
Surface gravity & $\log{g}$    & uniform &  & 3.0 to 6.0 cm/s$^2$ \\
Distance & $d$             & Gaussian &  & 21.14 $\pm$ 0.22 pc  \\
Flux scaling factor & $f$                   & uniform &  &  0.1 to 5.0 \\
Temperature at base of modeled atmosphere & $T_1$              & uniform &  &  1000 to 5000 K\\
$T$-$P$ profile coefficients$^{*}$ & $b_{i=1\dots6}$ & uniform &  & 0.1 to 0.95  \\
Error inflation term & $\ln{\delta}$   & uniform &  &  -10 to 1.0 \\
Mixing ratio of species i & $x_i$        & log-uniform &  &  $10^{-12}$ to 0.1 \\
\hline
\textit{gray clouds} & &  &  &   \\
Pressure at top of cloud & $p_{\mathrm{t}}$  & log-uniform &  & $10^{-2}$ to 50 bar  \\
Cloud base pressure scale factor & $b_{\mathrm{c}}$     & log-uniform &  &  1 to 10 \\
Optical depth & $\tau$              & uniform &  & -10 to 20 \\
\hline
\textit{non-gray clouds} & & &  &   \\
Pressure at top of cloud & $p_{\mathrm{t}}$  & log-uniform &  & $10^{-2}$ to 50 bar\\
Cloud base pressure scale factor & $b_{\mathrm{c}}$  & log-uniform &  &  1 to 10 \\
Optical depth at reference wavelength & $\tau_{\mathrm{ref}}$  & uniform &  & -10 to 20 \\
Particle size (dimensionless) with highest extinction & $Q_0$                 & log-uniform &  & 1 to 100  \\
Power-law index for small particles & $a_0$   & uniform &  &  3 to 7 \\
Particle size & $a$ & log-uniform &  & 0.1 to 50 $\SI{}{\micro\metre}$ \\
\hline
\end{tabular}
\begin{tablenotes}
\small
\item[*] The only, actually retrieved temperature $T$ is the one at the bottom of the modeled atmosphere. For all other temperatures, we retrieve a parameter $b_i$, such that, e.g., $T_2 = b_1\cdot T_1$.
\end{tablenotes}
\end{threeparttable}}
\end{table*}

For illustration, the spectra of both epochs were combined separately and their minimum and maximum values are displayed in Fig.~\ref{fig:HST18vsHST20}. It is apparent that the flux variation is wavelength-dependent.  It was suggested that heterogeneous clouds are the main driver for the observed spectral variability \citep{Bowler2020ApJ...893L..30B, Zhou2020AJ....160...77Z, Zhou2022AJ....164..239Z}.  Variability was found to be greater in 2018 (red) than 2020 (blue), when considering the flux ratios between maximum and minimum values of each epoch (see lower panel of Fig.~\ref{fig:HST18vsHST20}). 

As a complementary analysis, we considered the simultaneous $0.65-2.5 \mu$m wavelength-dependent medium-resolution ($3300 \leq$ R$_\lambda \leq 8100$) VLT/X-shooter spectrum published by \cite{Petrus2023A&A...670L...9P} (Fig.~\ref{fig:Xshooter_BF}) and the latest Early Release Science (ERS) JWST spectrum collected by \cite{Miles2023ApJ...946L...6M}, which observed VHS~1256~b with JWST's NIRSpec IFU and MIRI MRS modes, generating a spectrum spanning 1-18~$\mu$m at a $\sim$1000 - 3700 resolution (e.g., Fig.~\ref{fig:HELA_JWST_BF}).

\subsection{Bayesian Atmospheric Retrievals: \texttt{BeAR}}
\label{sect:BeAR}


We perform Bayesian atmospheric retrievals using tools that are widely adopted in past studies, e.g, it was previously applied to a curated sample of 19 L and T dwarfs \citep{Lueber2022ApJ...930..136L} as well as HIP~21152~B, the first T dwarf companion in the Hyades \citep{Franson2023AJ....165...39F}.

In this study we used the open-source Bern Atmospheric Retrieval code (BeAR). BeAR is a new and renamed version of the former Helios-r2 code described by \citet{Kitzmann2020ApJ} with enhanced capabilities and additional forward models. For this work we employed the emission spectroscopy forward model of BeAR. The code, together with a full documentation, has been released under a GNU General Public License v3.0 and can be found in a public GitHub repository\footnote{\url{https://github.com/newstrangeworlds/bear}}. The retrieval code uses the \texttt{MULTINEST} version of the Bayesian nested sampling \citep{Skilling2006, Feroz2009MNRAS} algorithm in order to explore the multi-dimensional parameter space of our models. The radiative transfer calculations are done via the method of short characteristics \citep{OlsonKunasz1987}.

The discretized temperature-pressure profile has 70 levels and is described by 7 parameters via the finite element approach \citep{Kitzmann2020ApJ}.  Bayesian model comparison follows standard approaches for calculating the Bayesian evidence and Bayes factor $B_{ij}$ \citep{Trotta2008ConPh}, which are natural outputs of the nested sampling algorithm \citep{Skilling2006}. A recognized association exists between the logarithm of the Bayes factor and the number of standard deviations by which one model is disfavored by the data  \citep{Trotta2008ConPh}.  Values of $\ln{B_{ij}}$~=~1, 2.5, and $>$5 indicate "weak", "moderate", and "strong" evidence for a statistical preference between the two models \citep{Trotta2008ConPh}. However, it is noteworthy that Bayesian model comparison may not always successfully rule out un-physical scenarios, as demonstrated by e.g. \cite{Fisher2019ApJ}.

The opacities (cross sections per unit mass) of atoms and molecules are computed using the open-source \texttt{HELIOS-K} calculator \citep{Grimm2015ApJ, Grimm2021ApJS} and publicly available via the \texttt{DACE} database \citep{Grimm2021ApJS}.  We include the following molecules: \ch{H2O}, \ch{CH4}, \ch{NH3}, \ch{CO2}, \ch{CO}, \ch{H2S}, \ch{CrH}, \ch{FeH}, \ch{CaH}, \ch{TiH}, \ch{TiO}, \ch{VO}, \ch{HCN}, as well as the alkali metals \ch{Na} and \ch{K}. The corresponding line lists are taken from the \texttt{ExoMol} database \citep{Barber2006MNRAS.368.1087B, Yurchenko2011MNRAS, Yurchenko2014, Azzam2016MNRAS} and the \texttt{HITEMP} database \citep{Rothman2010JQSRT}. Collision-induced absorption coefficients for \ch{H2}–\ch{H2} and \ch{H2}–\ch{He} are based on \citet{Abel2011JPCA} and \citet{Abel2012JChPh}, respectively. Following \citet{Kitzmann2020ApJ}, we use the line profile descriptions of \citet{Allard2016A&A} and \citet{Allard2019A&A} for the resonance lines of \ch{Na} and \ch{K}.

Our cloud model follows basic principles of Mie theory and was calibrated using Mie calculations based on measured refractive indices \citep{Kitzmann2018MNRAS}. This description of a non-gray cloud layer aims at parameterizing the particles' extinction coefficients. No attempt is made to actually model the formation of these clouds. The cross section of monodisperse cloud particles is designed to continuously transition from small to large particles, which correspond to a spectral slope and a constant (gray) cross section, respectively.  ``Small" and ``large" refer to the ``radiative" size of a particle, i.e., its physical size compared to the wavelength of radiation being absorbed or scattered.  This cloud model was implemented into \texttt{Helios-r2} in \cite{Lueber2022ApJ...930..136L}, and is now also part of the newest version of the code, which was renamed to \texttt{BeAR}.

All parameters and their prior distributions are shown in Tab.~\ref{tab:priors}. In total we have 22 free parameters for the cloud-free model, 25 free parameters for the gray cloud model and 28 for the non-gray cloud description, respectively.  Following \cite{Lueber2022ApJ...930..136L}, we adopt uniform prior distributions for the cloud optical depth, because that results in it being constrained. Note that we allow negative values in its prior range to ensure that the normal prior boundary of 0 is sampled correctly. Negative optical depths are internally replaced by 0 afterward to avoid un-physical results.

\subsection{Machine Learning Atmospheric Retrievals: \texttt{HELA}}
\label{sect:HELA}

The use of supervised machine learning allows one to perform approximate Bayesian computation using pre-computed model grids.  In our open-source \texttt{HELA} computer code, we implemented the random forest method for performing atmospheric retrievals \citep{Marquez2018NatAs...2..719M}, which was subsequently used in several follow-up studies on analyzing high-resolution spectra of the ultra-hot Jupiter KELT-9b \citep{Fisher2020AJ....159..192F}, medium-resolution spectra of benchmark brown dwarfs \citep{Oreshenko2020AJ, Lueber2023ApJ...954...22L}, the information content of \textit{James Webb} Space Telescope spectra (JWST; \citealt{Guzman2020AJ....160...15G}) and the sampling strategy of model grids \citep{Fisher2022ApJ...934...31F}.  Training and testing utilize 80\% and 20\% of the training set, respectively, as was established in these earlier studies.

We select three legacy model grids of brown dwarf atmospheres to use as training sets for \texttt{HELA}.  The model grid of \cite{Hubeny2007ApJ...669.1248H}, computed using the \texttt{COOLTLUSTY} code \citep{Sudarsky2003ApJ...588.1121S, Hubeny2003ApJ...594.1011H, Burrows2006ApJ...640.1063B}, includes both equilibrium and disequilibrium chemistry for cloud-free and cloudy atmospheres.  For analyzing the spectra of VHS~1256~b, we utilize only the cloudy models with disequilibrium chemistry. This disequilibrium chemistry is represented by vertical mixing, characterized by a vertical eddy diffusion coefficient K$_{zz}$, which is set to 10$^{6}$ cm$^2$s$^{-1}$ in the current study. The model grid covers $4.5 \le \log{g} \le 5.5$ (cgs units) and $700 \le T_{\rm eff}/\mbox{K} \le 1800$ in steps of $\Delta \log{g}=0.5$ and $\Delta T_\mathrm{eff}=100$ K, respectively. To improve the predictability of model parameters, we perform a one-dimensional linear interpolation of the measured flux for sufficiently small surface gravity steps ($\Delta \log{g}=0.025$) to increase the total grid size from 208 to 480 spectra (see discussion in \citealt{Lueber2023ApJ...954...22L}).

The \texttt{BT-Settl} model grid \citep{Allard2012EAS....57....3A} was computed using the \texttt{PHOENIX} computer code \citep{Hauschildt1992JQSRT..47..433H, Hauschildt1997ApJ...488..428H}.  It accounts for vertical convective mixing and overshooting \citep{Allard2011ASPC..448...91A}.  The \texttt{BT-Settl} model grid covers $2.0 \le \log{g} \le 5.5$ (cgs units) and $500 \le T_{\rm eff}/\mbox{K} \le 2400$ in steps of $\Delta \log{g}=0.5$ and $\Delta T_\mathrm{eff}=50-100$ K, respectively. Again, we perform a one-dimensional linear interpolation of the measured flux for surface gravity ($\Delta \log{g}=0.05$), resulting in an increase of the total grid size from 216 to 1237 spectra.

The most modern model grid among the three sets is the cloudy \texttt{Sonora Diamondback} model \citep{Morley2024arXiv240200758M}, which covers $3.5 \le \log{g} \le 5.5$ (cgs units) and $900 \le T_{\rm eff}/\mbox{K} \le 2400$ in steps of $\Delta \log{g}=0.50$ and $\Delta T_\mathrm{eff}=100$ K, respectively. Based on the radiative-convective equilibrium model by \cite{Marley1999Icar..138..268M}, chemical equilibrium holds throughout the atmosphere. The model accounts for vertical mixing using mixing length theory within the \cite{AckermanMarley2001ApJ...556..872A} cloud parametrisation. Three metallicities [M/H] are available (solar, +0.5, -0.5), as well as cloud parameter $f_\mathrm{sed}$ from 1.0 to 8.0. The one-dimensional linear interpolation of the measured flux was performed to achieve surface gravity steps of $\Delta \log{g}=0.025$ cm/s$^2$ and a total grid size of 11521 individual spectra, compared to 1440 in the original grid.

In addition to the mentioned varying parameters of our considered grids, we introduce an additional parameter, a so-called flux scaling factor $f$, to our retrievals. Its purpose lies in scaling the radius–distance relationship for the outgoing flux $F_{\nu}^{+}$ of the brown dwarf to the one measured by the observer ($F_{\nu}$):
\begin{equation}\label{eq:radius-distance relation}
    \centering
    F_{\nu}=F_{\nu}^+ f\left(\frac{\mathcal{R}}{d}\right)^2 \, ,
\end{equation}
where $d$ is the distance between the observer and the brown dwarf and $\mathcal{R}$ the prior brown dwarf radius. To incorporate it into our training sets, we assume a uniform distribution of $0.5 \leq f \leq 2.0$. We randomly draw an $f$ for each model spectrum and multiply the entire set of fluxes by it.

\section{Results}
\label{sect:Results}

\subsection{Standard Bayesian retrieval analysis and sensitivity test of retrieved radius using HST data}
\label{subsect:standard}

\begin{figure}
    \centering
    \includegraphics[width=\columnwidth]{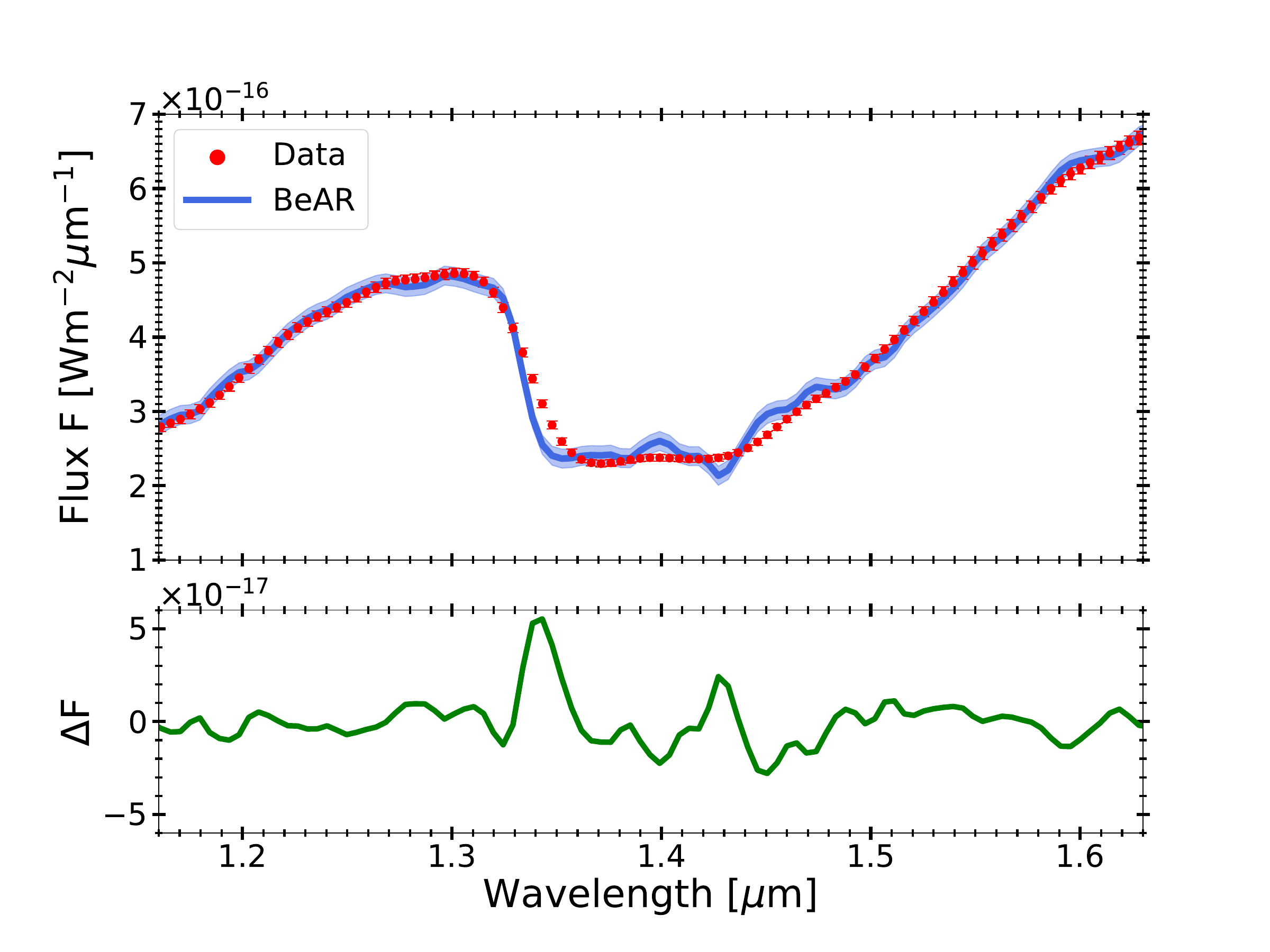}
    \caption{Posterior median spectra (F) and residuals ($\Delta$ F) associated with the free-chemistry retrieval analyses of the HST 2018 maximum brightness spectrum with a gray-cloud model. Data are shown as red dots with associated uncertainties.}
\label{fig:VHS_retrieval_HST18max_gray_spectra}
\end{figure}

\begin{figure*}[ht!]
    \centering
    \includegraphics[width=0.99\columnwidth]{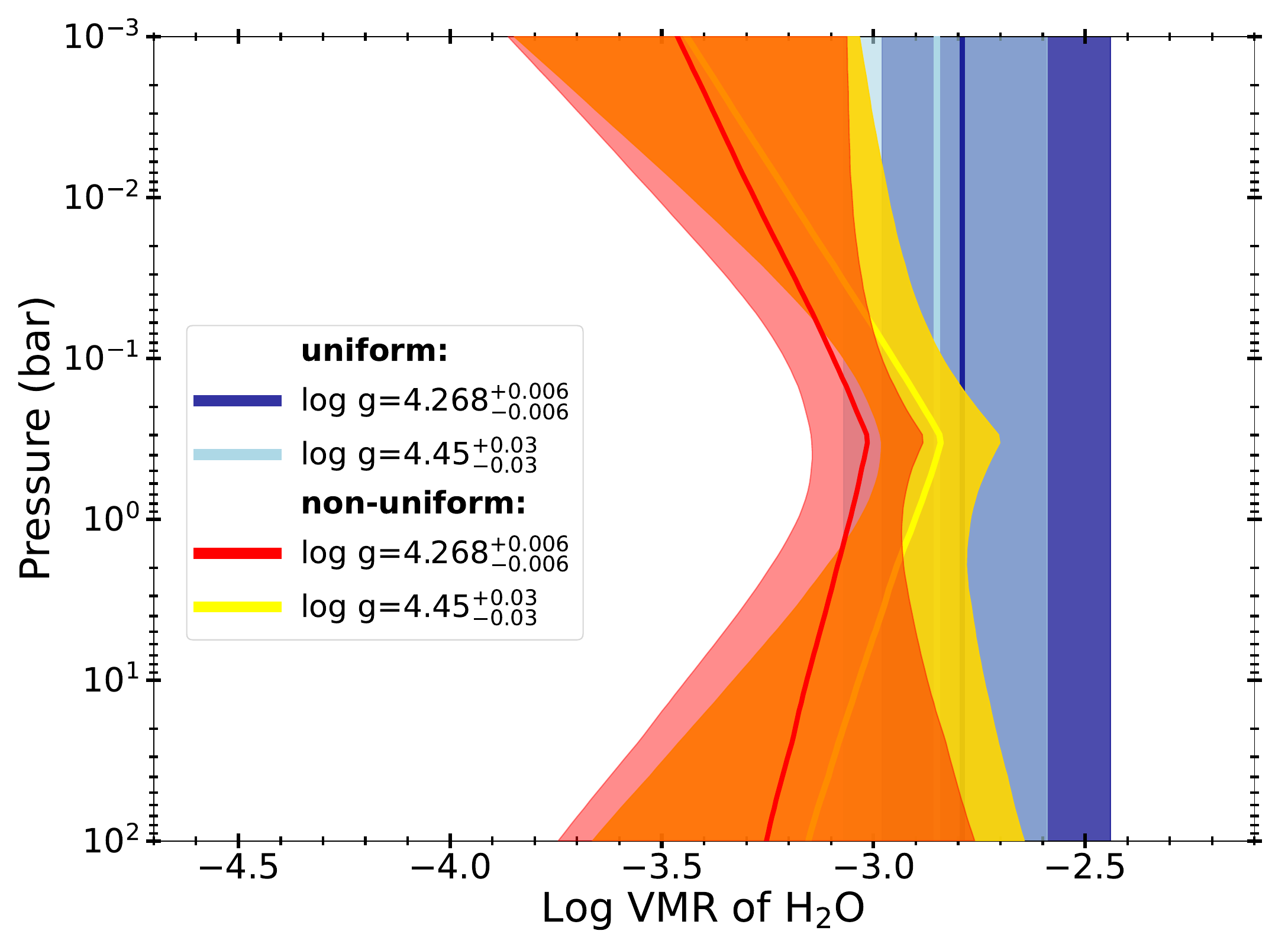}
    \includegraphics[width=\columnwidth]{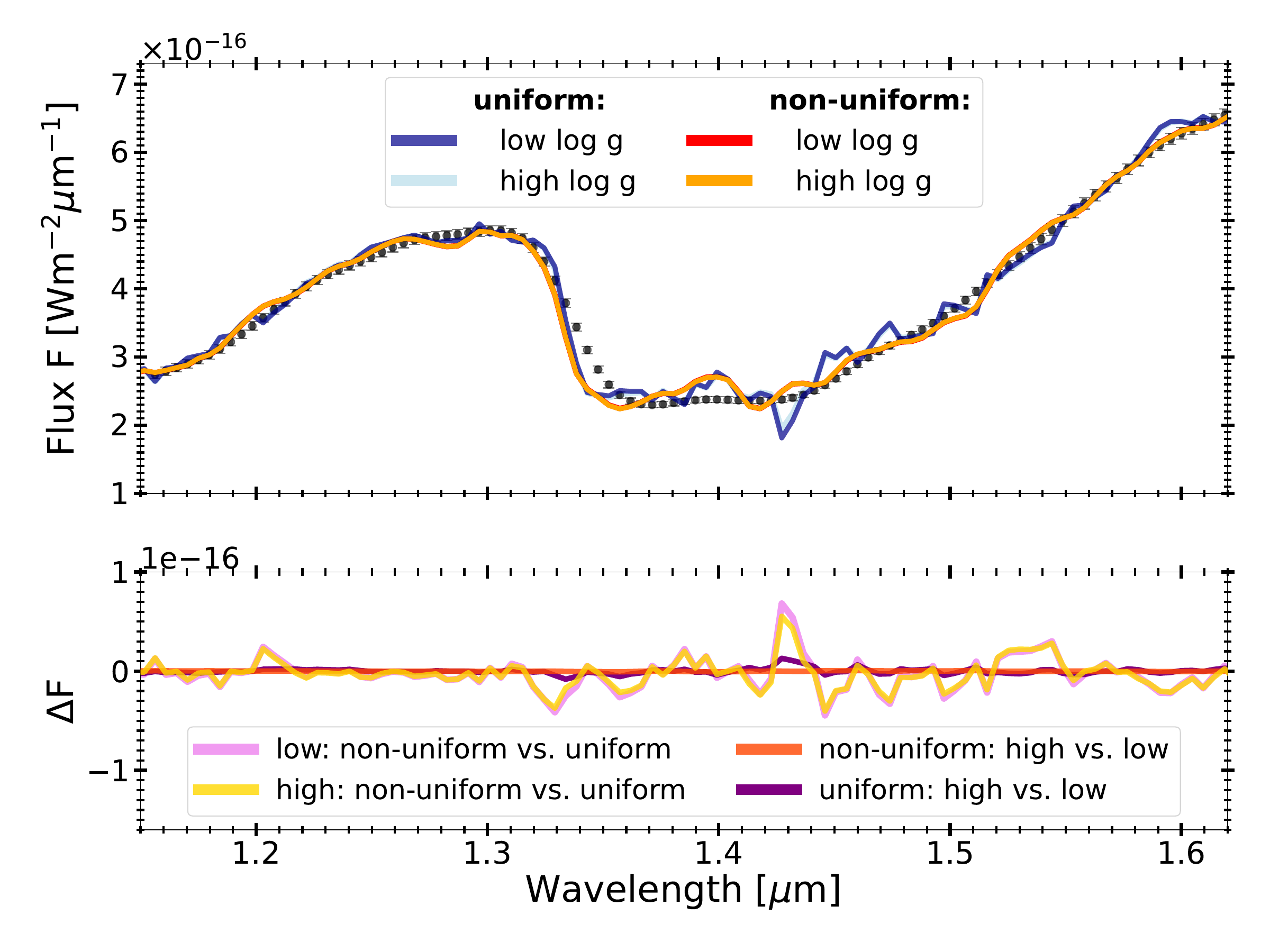}
    \caption{Left: Retrieved volume mixing ratios for \ch{H2O} of the HST 2018 maximum brightness spectrum. Four retrieval runs were conducted: a vertically uniform and a non-uniform \ch{H2O} abundance profile one for each of the two fixed surface gravity values of $\log{g}=4.268 \pm 0.006$ (low) and $\log{g}=4.45 \pm 0.03$ (high). Right: Corresponding posterior median spectra (F) for all four runs and differences between them ($\Delta$ F). The HST 2018 maximum brightness spectrum data is shown as black dots with associated uncertainties.}
    \label{fig:H2Oabdundance}
\end{figure*}

We perform a suite of HST retrievals using the \texttt{BeAR} code, applied on the minimum, average and maximum spectra (in terms of the flux) from each epoch using models without clouds, with gray clouds and with non-gray clouds.  The outcome of this suite of retrievals is reported in Tab.~\ref{tab:data posteriors} in terms of the retrieved atmospheric properties.  Generally, the quantities that are meaningfully constrained are the surface gravity, photometric radius, water abundance (by volume), cloud optical depth and cloud-top pressure. Tab.~\ref{tab:Bayes data} compares the Bayesian evidences and natural logarithm of the Bayes factors $\ln B_{ij}$, which demonstrates that cloudy models are strongly favored \citep{Trotta2008ConPh}.  Additionally, the Bayesian model comparison favors a gray cloud description over non-gray cloudy models. Therefore, Figs.~\ref{fig:VHS_retrieval_HST18max_gray_spectra} and \ref{fig:VHS_retrieval_HST18max_gray} show the best-fit model outcome of our HST atmospheric retrieval analysis for the HST 2018 maximum brightness spectrum with a gray cloud description, including the joint posterior distributions of the parameters as well as the median spectra.  The retrieved posterior distributions are for the surface gravity ($\log{g}$), the radius ($R$; as explained in Eq. [10] of \citealt{Kitzmann2020ApJ}), the distance to VHS~1256~b ($d$), and the chemical abundances and the cloud properties. The cloud properties include the optical depth $\tau$, the cloud-top pressure $p_t$, the cloud-bottom pressure $p_b = b_c p_t$, the (spherical) cloud particle radius $a$, the spectral slope associated with small particles $a_0$, a proxy for the particle composition $Q_0$ and the reference optical depth $\tau_{\rm ref}$ at 1 $\mu$m.  As we are displaying the exemplary outcome for the HST 2018 maximum brightness spectrum retrieval with gray clouds, we have only shown the posterior distributions of $\tau$, $p_t$ and $p_b$. Except for water, the chemical compositions of molecules are poorly constrained or unconstrained.

The retrieved surface gravity values of $\log{g} \sim$ 5.2--5.7, are reasonable for field/high-mass brown dwarfs, but certainly higher than expected by previous studies of \cite{Gauza2015ApJ...804...96G} ($\log{g}$ = 4.24$\,\pm\,$0.35) or \cite{Miles2018ApJ...869...18M} ($\log{g}$ = 3.2), which are based on model grids. The retrieved effective temperatures $T_\mathrm{eff} \sim$ 1360--1390 K are approximately 100-150 K hotter than the value obtained by \cite{Miles2018ApJ...869...18M} with 1240 K. The prediction by \cite{Gauza2015ApJ...804...96G} with T$_\mathrm{eff}$=$880^{+140}_{-110}$ is about 400–700 K cooler than the temperature expected for late-L field dwarfs. Retrieved radii R~$\sim$~0.7--0.8~R$_{\rm J}$ are lower than expected given the age of VHS~1256~b. Similarly small radii have been inferred in numerous other retrieval studies (e.g., \citealt{Zalesky2019ApJ, Gonzales2020ApJ, Lueber2022ApJ...930..136L}). \cite{Dupuy2023MNRAS.519.1688D} derived estimates from using the hybrid cloudy evolutionary model by \cite{SaumonMarley2008ApJ} suggesting two approximate radii R of $1.30\,R_{\rm Jup}$ and $1.22\,R_{\rm Jup}$, effective temperatures $T_{\rm eff}$ of $1153\pm5\,$K and $1194\pm9\,$K, as well as surface gravities of $\log{g}=4.268 \pm 0.006$ and $\log{g}=4.45 \pm 0.03$, according to the two potential mass estimates for VHS~1256~b of 12.0$\,\pm\,$0.1~M$_\mathrm{Jup}$ and 16$\,\pm\,$1~M$_\mathrm{Jup}$.

A closer look at the water abundances retrieved from the HST spectra with \texttt{BeAR} reveals somewhat high values and a strong positive correlation with surface gravity (see Fig.~\ref{fig:VHS_retrieval_HST18max_gray} and Tab.~\ref{tab:data posteriors}). These results align with the findings of \cite{Phillips2024ApJ...972..172P}, who also reported somewhat high ($\sim$1\% to 10\%) H$_2$O abundances for two late-L objects, W0047 and BD+60 1417B, using SpeX data. This emphasizes the challenges associated with characterizing water in low-gravity L dwarfs. After performing several tests, which included fixing the gravity based on values from previous publications or adjusting the pressure range of the atmosphere, we conclude that water and surface gravity show a degeneracy in these atmospheric retrievals on HST spectra since the \ch{H2O}-content depends strongly on the assumed prior of $\log{g}$. It should be noted that these adjustments did not affect the retrieved effective temperatures, radii, or cloud parameters. Thus, for our further runs, we decided to fix the prior of $\log{g}$ to the two values taken from \cite{Dupuy2023MNRAS.519.1688D}: $\log{g}=4.268 \pm 0.006$ and $\log{g}=4.45 \pm 0.03$, respectively.  Since these values are derived from evolutionary models, our approach is somewhat circular.  We later note that the derived radius and effective temperatures are markedly different from the ones corresponding to the evolutionary models used by \cite{Dupuy2023MNRAS.519.1688D}.

As an additional test, we conduced retrieval test runs with a vertically non-uniform \ch{H2O} abundance profile for the HST 2018 maximum brightness spectrum, for both fixed surface gravity values. For the vertically non-uniform description of the \ch{H2O} abundance, \texttt{BeAR} implements a description of the abundance profile based on a finite-element approach, analogous to the temperature profile described in Eq. [12] of \cite{Kitzmann2020ApJ}. Unless stated otherwise, we use two first-order elements for the abundance profile, which results in a total of three free additional parameters for the retrieval. Comparing the retrieval statistics of the vertically non-uniform water abundance models against the corresponding uniform models results in a strong preference for vertically varying \ch{H2O} abundance profiles ($\ln{B_{ij}\approx13.4}$ for a fixed $\log{g} = 4.268 \pm 0.006$ and $\ln{B_{ij}\approx 13.1}$ for $\log{g} = 4.45 \pm 0.03$, respectively). However, Fig.~\ref{fig:H2Oabdundance} also graphically highlights that both abundance profile descriptions agree in the range of greatest contribution (0.1--1 bars, see Fig.~\ref{fig:VHS_retrieval_HST18max_gray}) within the respective uncertainties. The non-constant vertical water profiles may arise from latitudinal or longitudinal inhomogeneities within the brown dwarf atmosphere, manifesting in non-constant profiles in a 1D atmospheric retrieval.

\subsection{Time series HST retrieval analysis}
\label{subsect:time_series}

\begin{figure*}[ht!]
    \centering
    \includegraphics[width=\columnwidth]{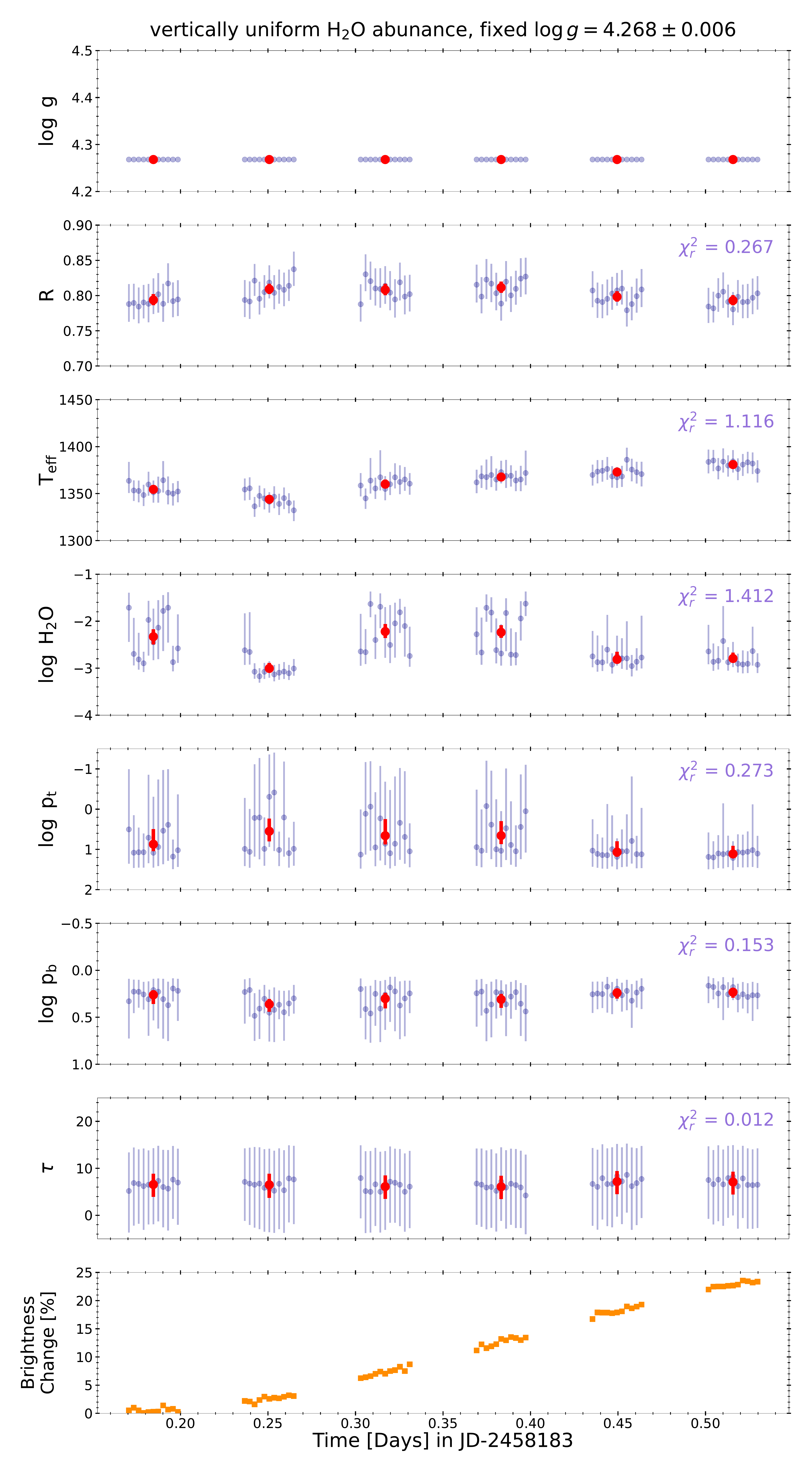}
    \includegraphics[width=\columnwidth]{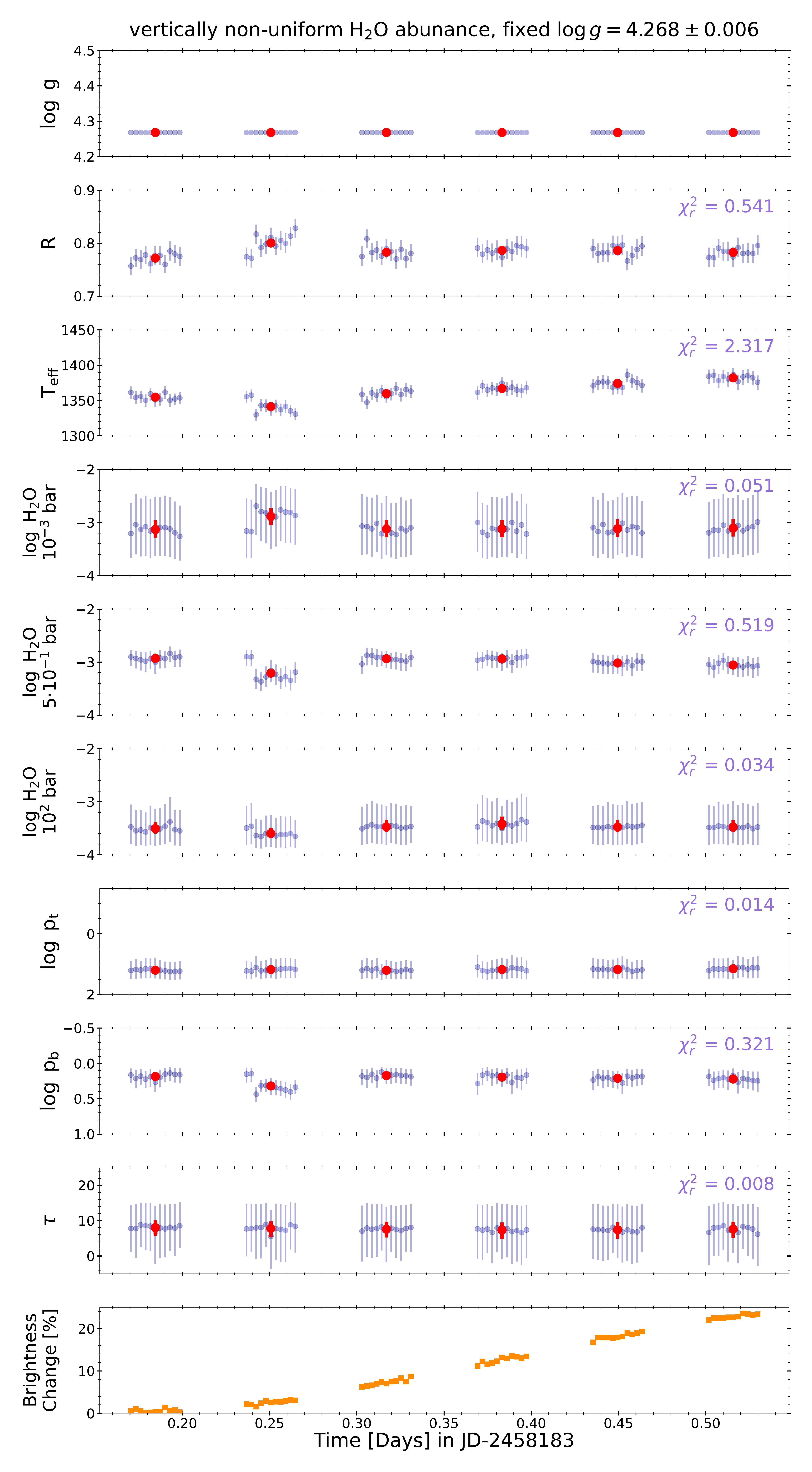}
    \caption{HST18 \texttt{BeAR} posteriors with a fixed surface gravity of $\log{g}=4.268 \pm 0.006$ and vertically uniform (left column) vs. non-uniform (right column) \ch{H2O} abundances as a sequence over all 6 orbits in 2018. Each orbit consist of 11 intra-orbit spectra. Orbit-averaged posteriors are shown in red. The reduced chi-square values correspond to a fit of a straight line through all posteriors of a given parameter. The lowest panel represents the brightness change in G141 broad band, normalized by $3.0 \times 10^{-13}$~erg/s/cm$^2$/$\mu$m (data taken from \citealt{Bowler2020ApJ...893L..30B} and \citealt{Zhou2022AJ....164..239Z}).}
    \label{fig:Comparison_GN_Chi2}
\end{figure*}

Next, we apply our retrieval framework towards analyzing an entire time series of spectra.  Since data from the 2018 epoch displays the most variability, we choose to focus our efforts on the 66 spectra collected during the same HST visit. Following the outcome of the previous sub-section, we use only cloudy models with gray clouds as well as both vertical uniform water and non-uniform \ch{H2O}-abundance profiles. Fig.~\ref{fig:Comparison_GN_Chi2} shows the retrieved atmospheric properties across the 66 different spectra, which are constructed from 6 different orbits of VHS~1256~b with 11 spectra in each orbit. We fixed the surface gravity to a value of $\log{g}=4.268 \pm 0.006$ due the degeneracy with the water abundance. The analogous plot for a surface gravity value of $\log{g} = 4.45 \pm 0.03$ can be found in Fig.~\ref{fig:Comparison_GN_445_Chi2}.

We note that the retrieved radius ($\approx 0.8~R_{\rm J}$) and effective temperatures ($\approx 1350$ K) are markedly different compared to the values associated with the evolutionary models used by \cite{Dupuy2023MNRAS.519.1688D}, who reported $R \approx1.2$--$1.3~R_{\rm J}$ and $T_{\rm eff} \approx 1200$ K.  In other words, despite using the surface gravities reported by \cite{Dupuy2023MNRAS.519.1688D} as priors for our retrievals the retrieved radii and effective temperatures are discrepant.  This suggests that different input physics and chemistry were used in the analyses and highlights their model-dependent nature.

By fitting a flat line to the retrieved quantities across time, both for vertical uniform non-uniform description of \ch{H2O}, we find that almost all of the quantities result in reduced chi-square values well below unity. This implies the existence of two potential scenarios: either the quantities remain invariant over time, or our observations and retrievals lack the sensitivity required to detect any temporal variability. As expected from previous studies, the effective temperature does increase with time and is highly correlated with the variability amplitude of VHS~1256~b, resulting in Pearson correlation coefficients of $\approx$ 0.85. The only exception next to the effective temperature slightly increasing with time is the abundance of \ch{H2O}, which exhibits a slight variation between the first, second and third orbits. The variation over time is limited to a pressure of around 0.1--1 bars in the vertically non-uniform description, corresponding to the pressure range in which the largest contribution comes from. 

The temporal variability of VHS~1256~b cannot be attributed to a single varying factor, probably because the various atmospheric properties are degenerate with one another.  It is tempting to conclude that the atmospheric properties adjust rapidly to seek an equilibrium and therefore a stable climate.  But without spectra covering a wider wavelength range, at different epochs, it is premature to draw this conclusion.

\begin{figure*}
  \centering
  \includegraphics[width=.33\textwidth]{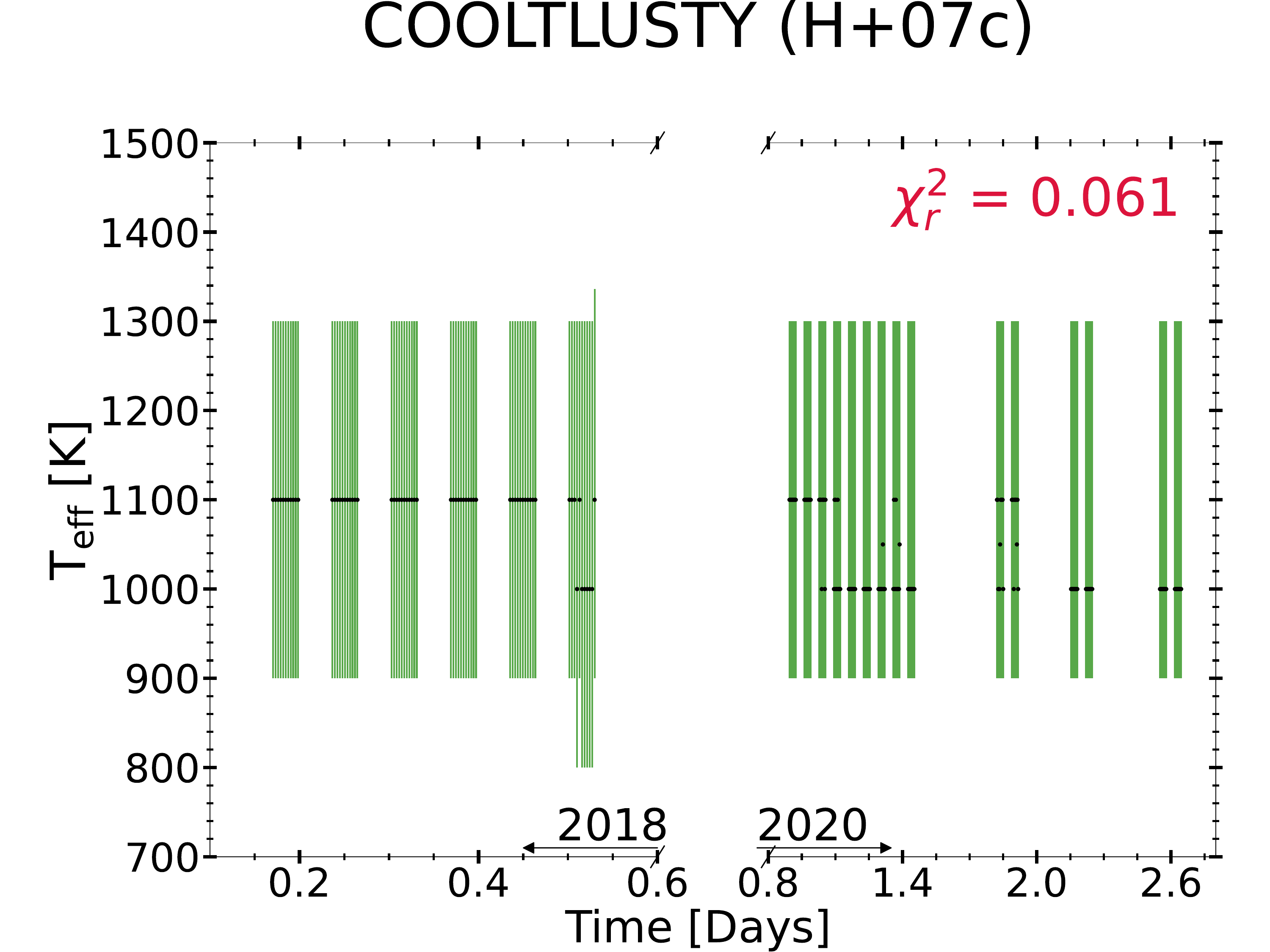}\hfill
  \includegraphics[width=.33\textwidth]{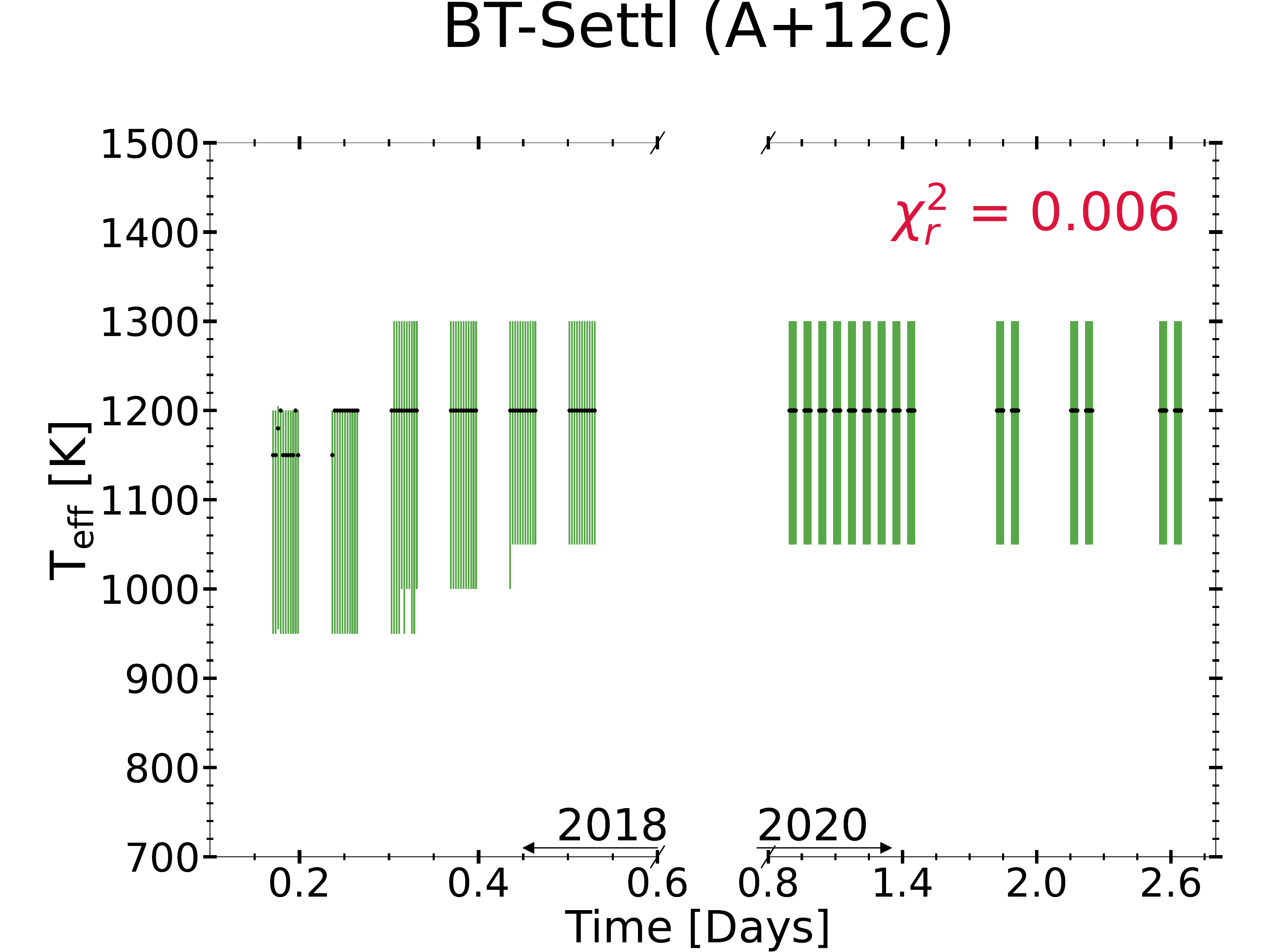}\hfill
  \includegraphics[width=.33\textwidth]{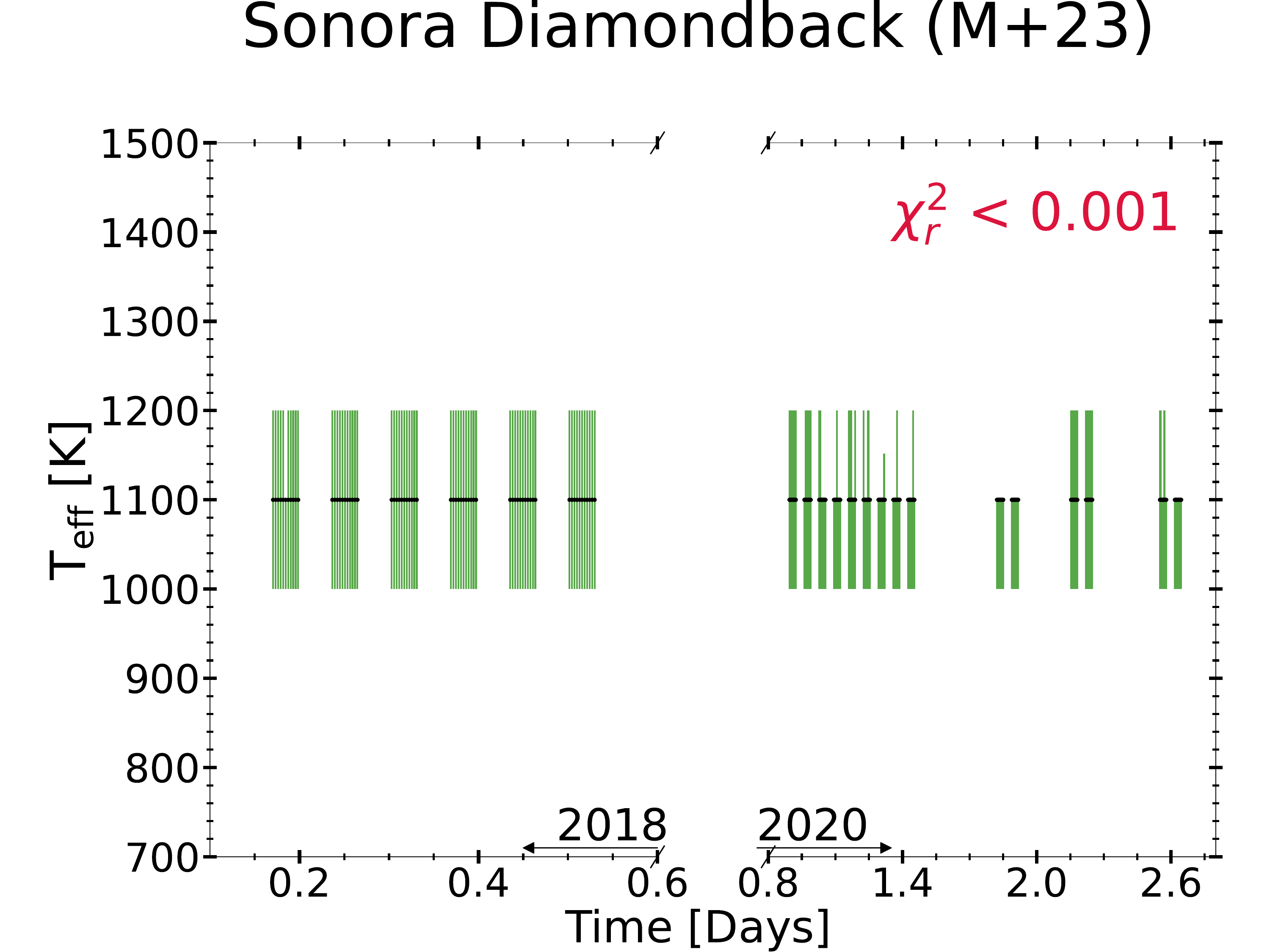}\hfill
  \includegraphics[width=.33\textwidth]{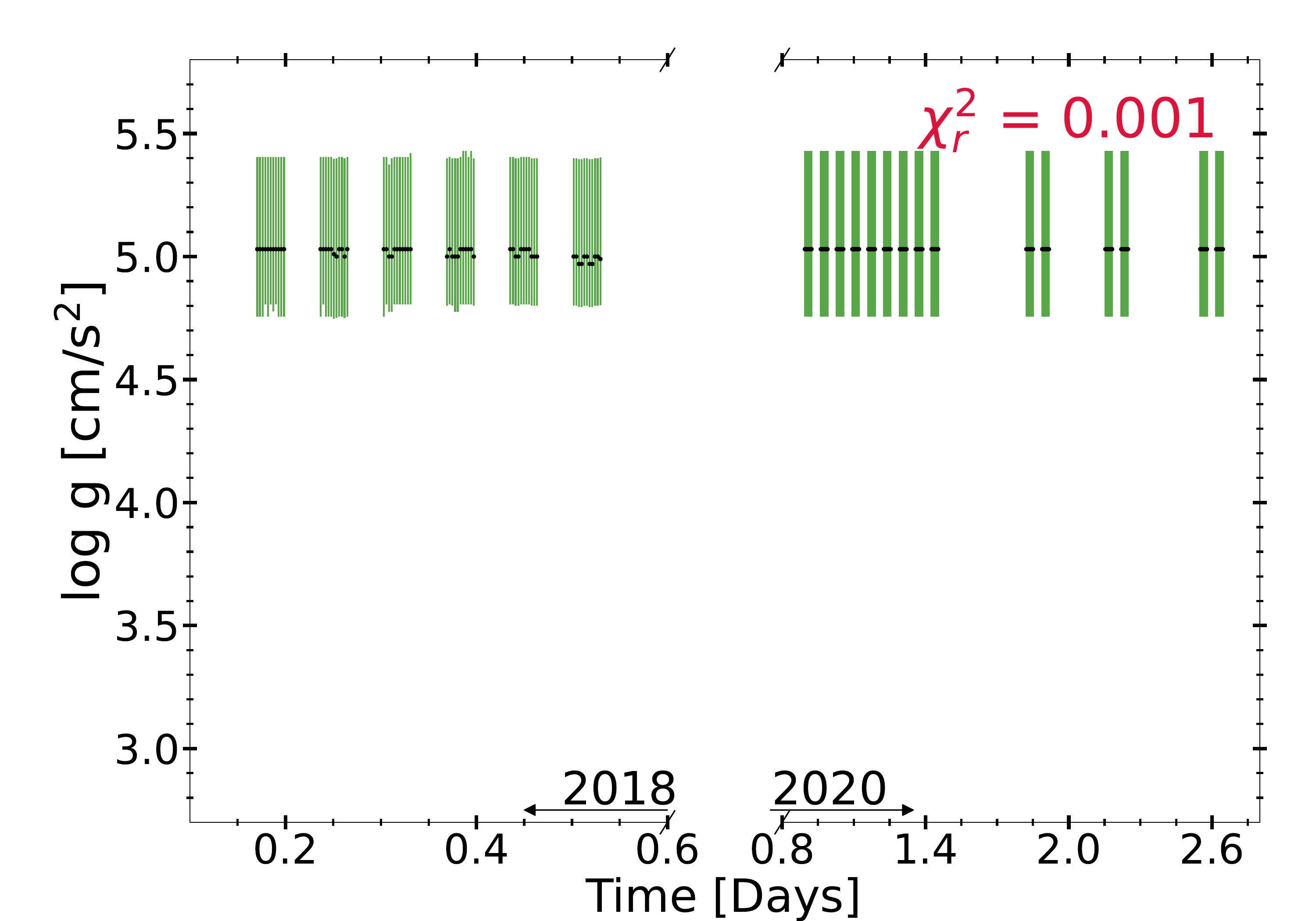}\hfill
  \includegraphics[width=.33\textwidth]{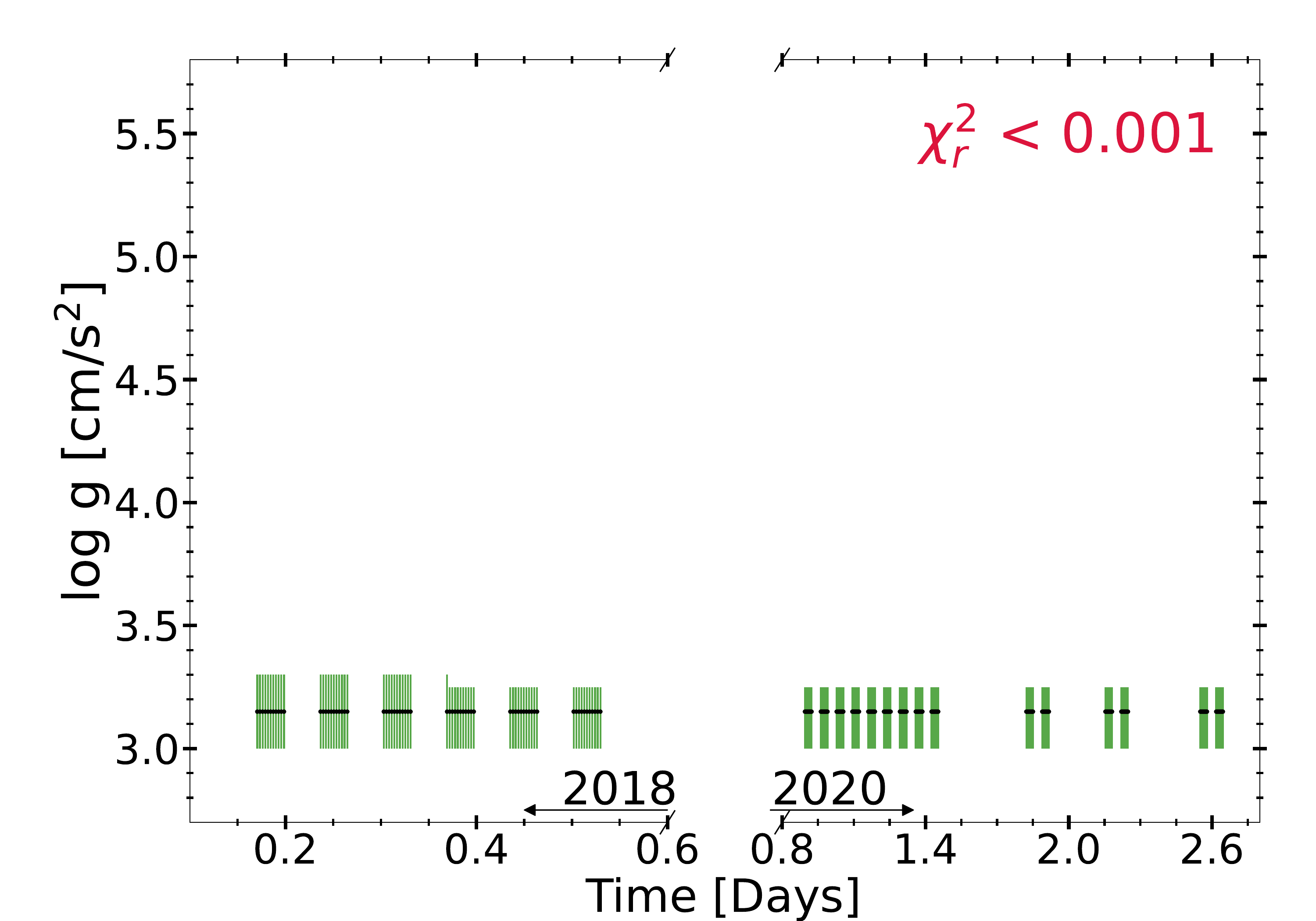}\hfill
  \includegraphics[width=.33\textwidth]{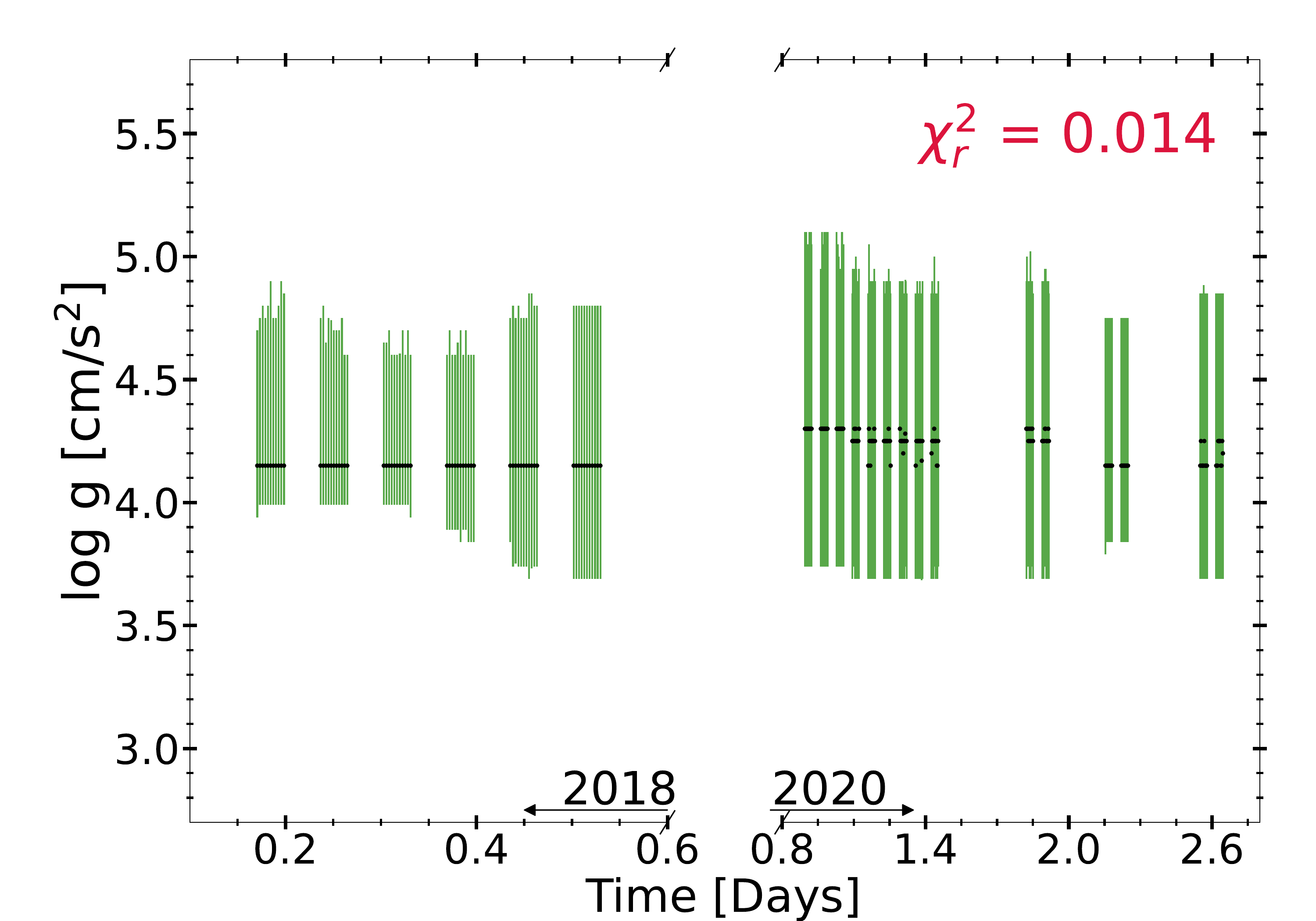}\hfill
  \includegraphics[width=.33\textwidth]{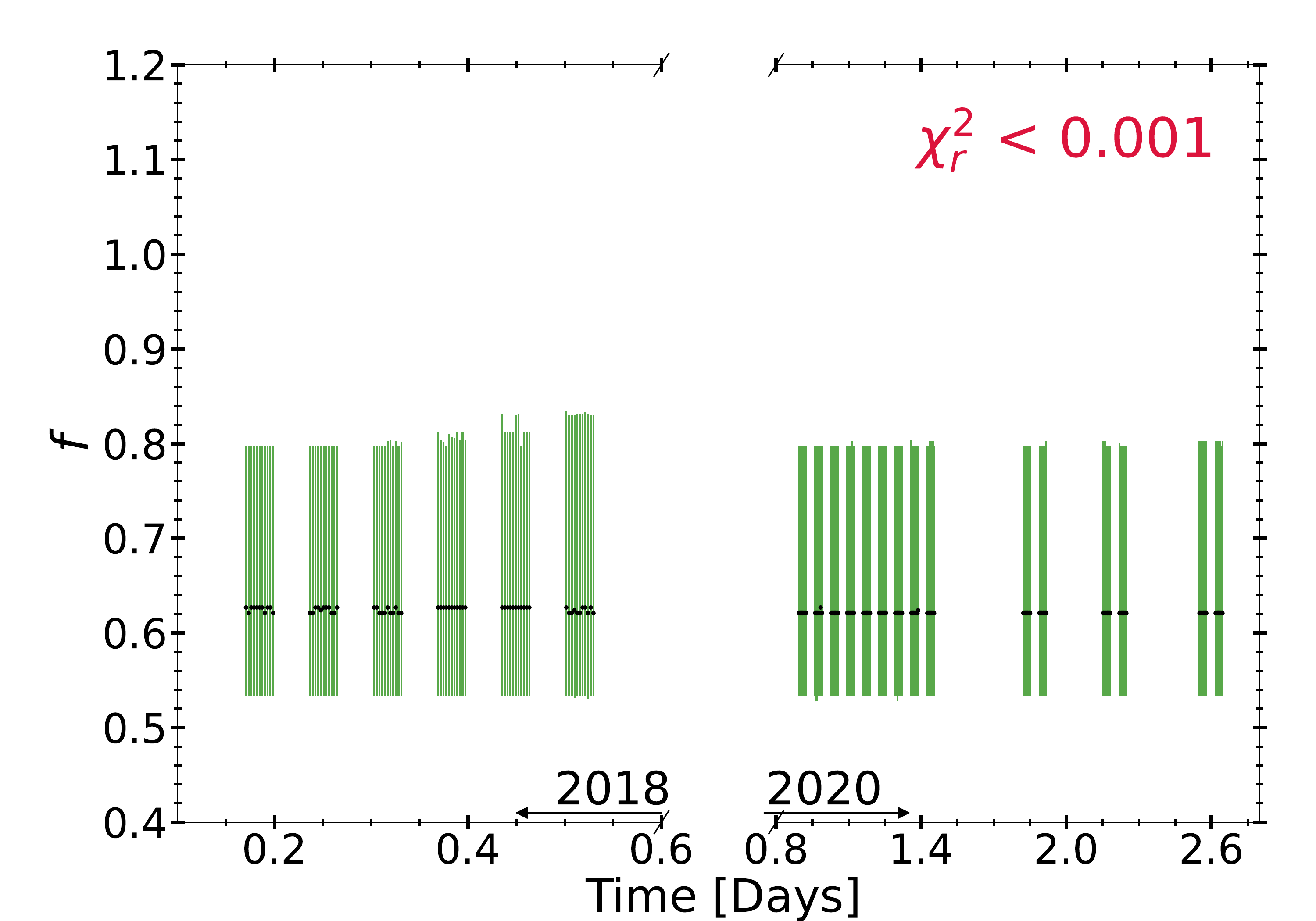}\hfill
  \includegraphics[width=.33\textwidth]{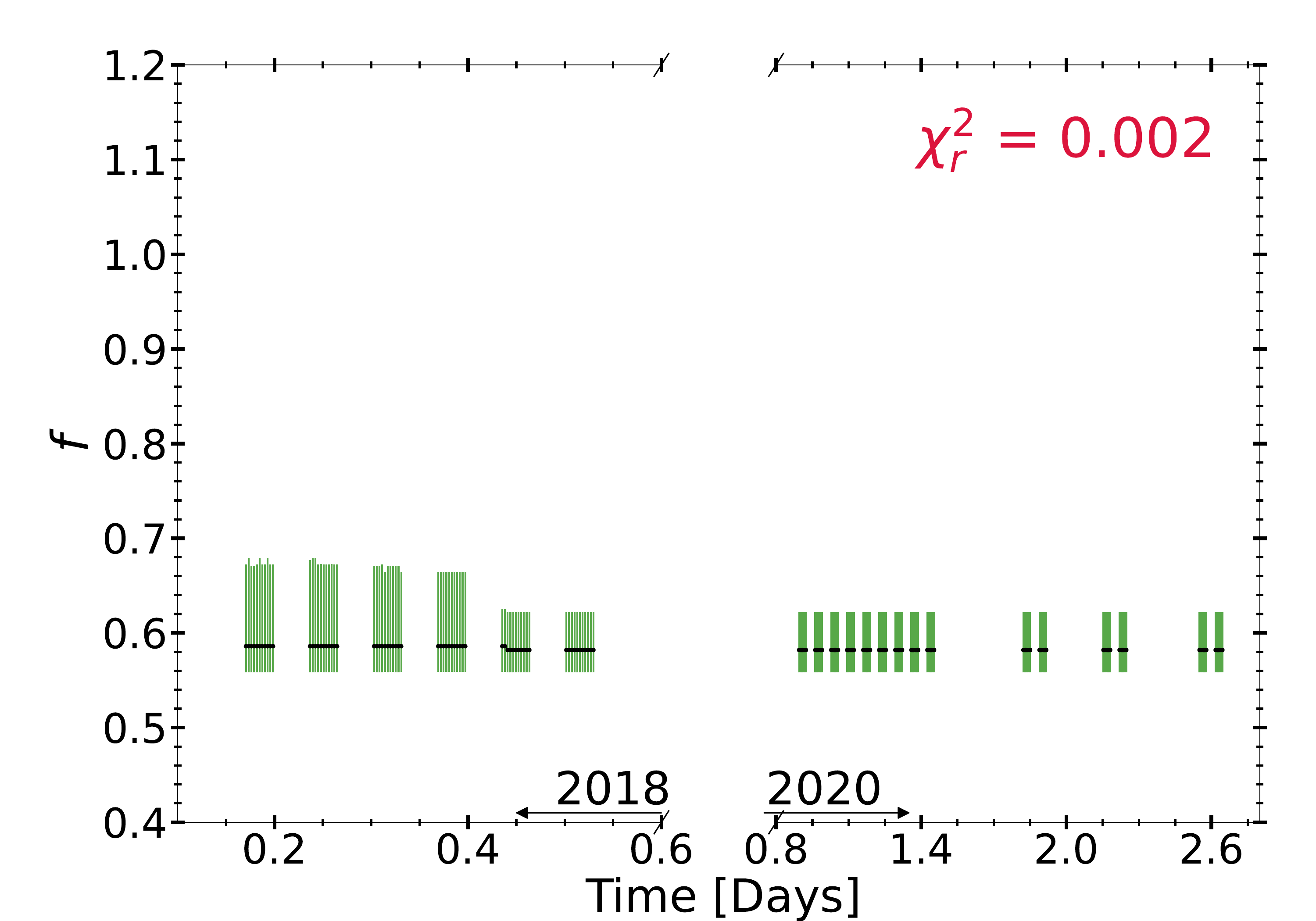}\hfill
  \includegraphics[width=.33\textwidth]{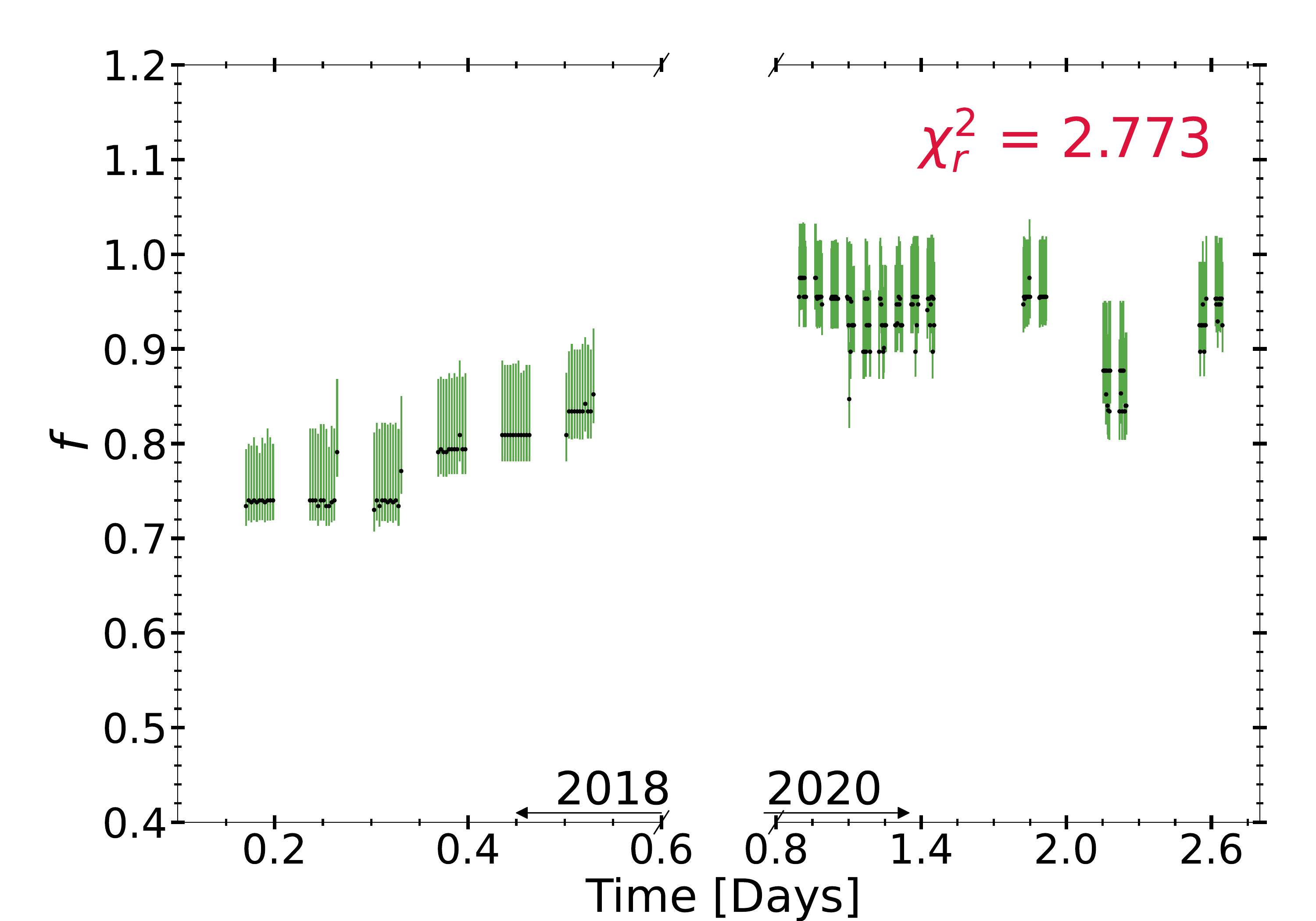}
  \caption{Machine learning retrieval outputs for legacy model grids. From left to right: \texttt{COOLTLUSTY} from \cite{Hubeny2007ApJ...669.1248H}, \texttt{BT-Settl} by \cite{Allard2012EAS....57....3A}, and \texttt{Sonora Diamondback} by \cite{Morley2024arXiv240200758M}. The HST exposures cover data from 2018 (66) and 2020 (165), where the x-axes are broken to fit the two epochs of data. Each orbit consist of 11 intra-orbit spectra. Posterior median values are indicated as black dots, while corresponding error bars are shown in green. The represented reduced chi-square values correspond to a fit of a straight line through all posteriors of a given parameter.}
  \label{fig:ML_grids}
\end{figure*}

\begin{figure*}
  \centering
  \includegraphics[width=.33\textwidth]{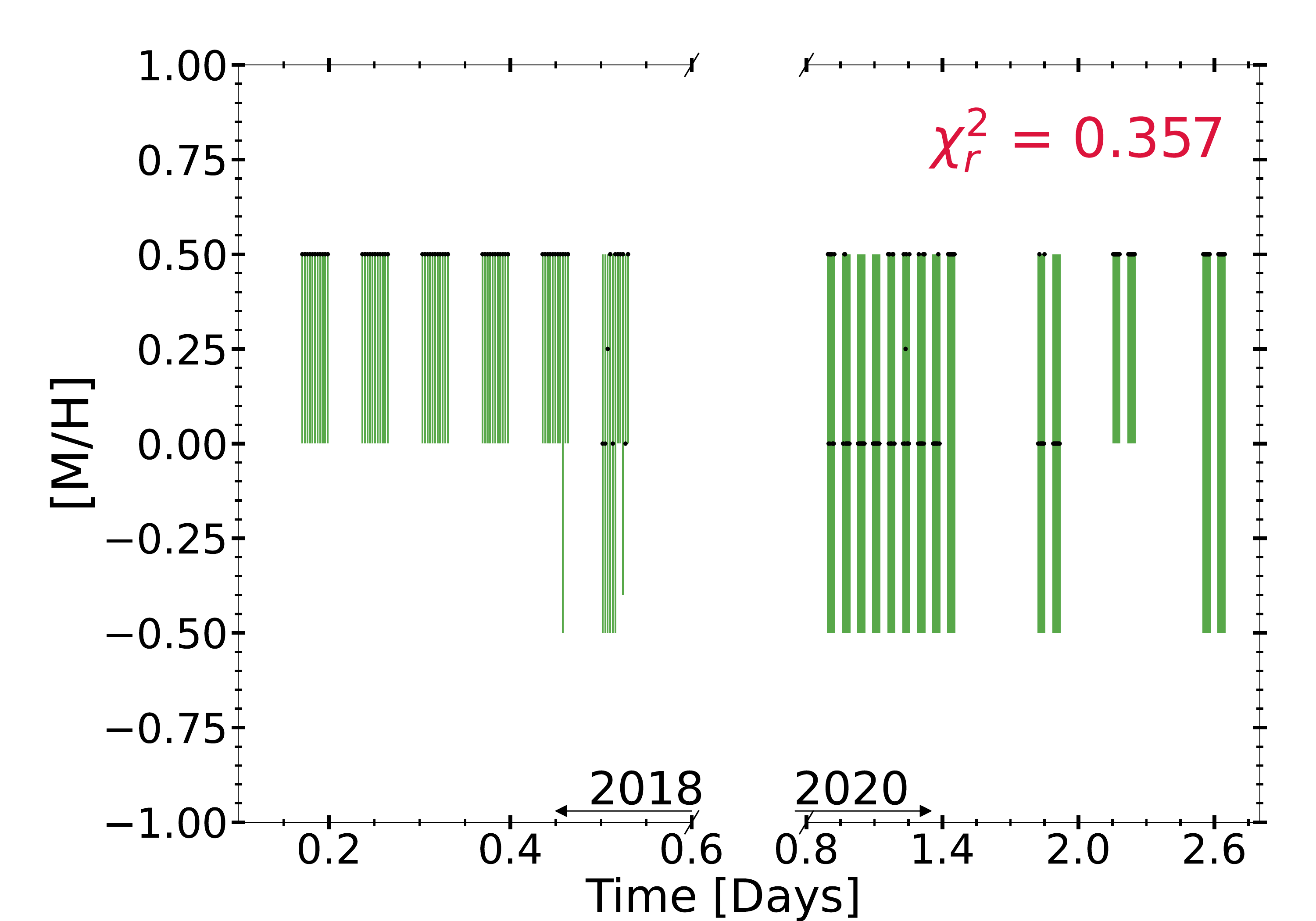}
  \includegraphics[width=.33\textwidth]{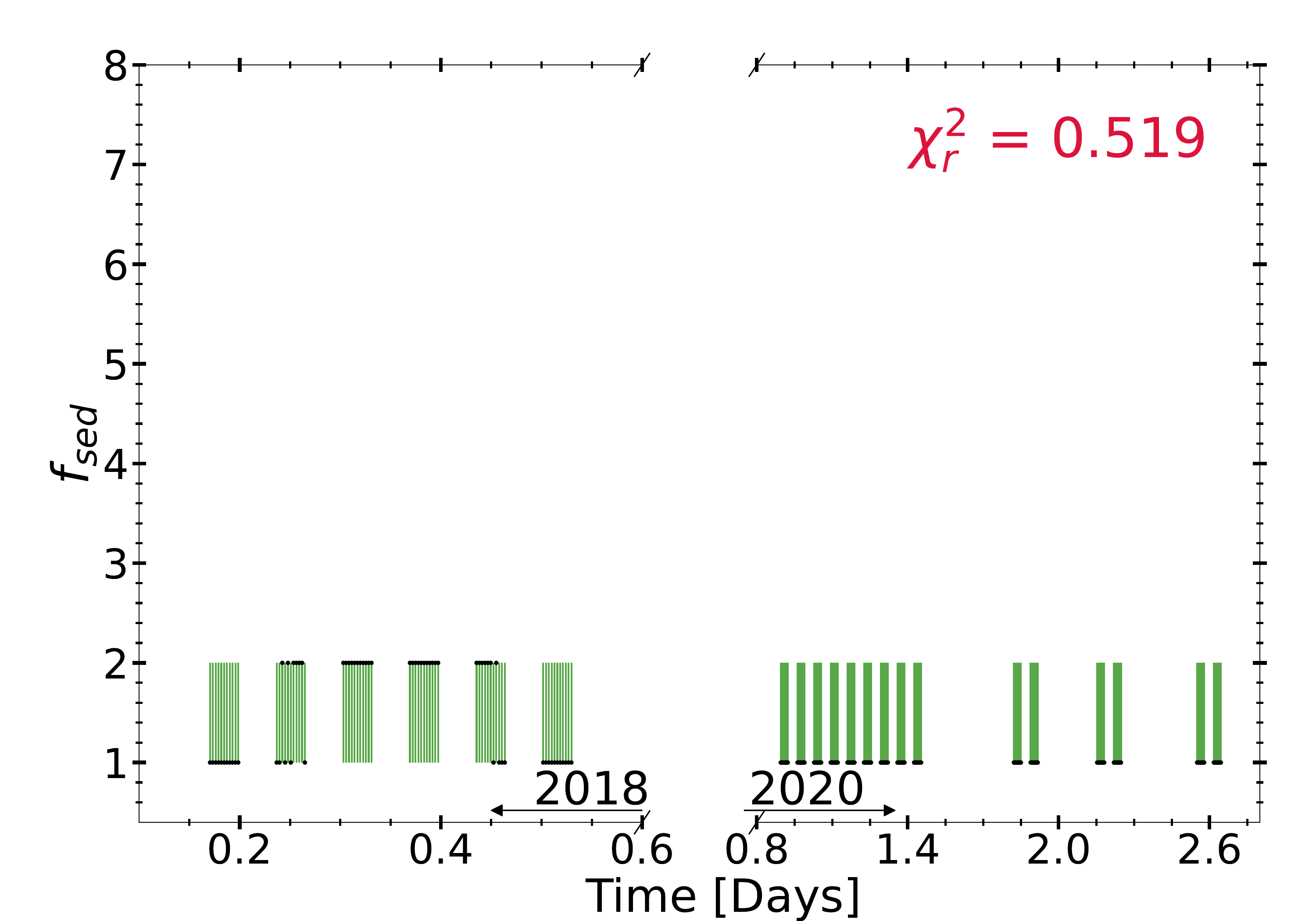}
  \caption{Additional machine learning retrieval outputs for legacy model grid \texttt{Sonora Diamondback} by \cite{Morley2024arXiv240200758M}. The HST exposures cover data from 2018 (66) and 2020 (165), where the x-axes are broken to fit the two epochs of data. Each orbit consist of 11 intra-orbit spectra. Posterior median values are indicated as black dots, while corresponding error bars are shown in green. The represented reduced chi-square values correspond to a fit of a straight line trough all posteriors of a given parameter.}
  \label{fig:ML_grids_add}
\end{figure*}

\begin{figure*}[ht!]
    \centering
    \includegraphics[width=0.75\textwidth]{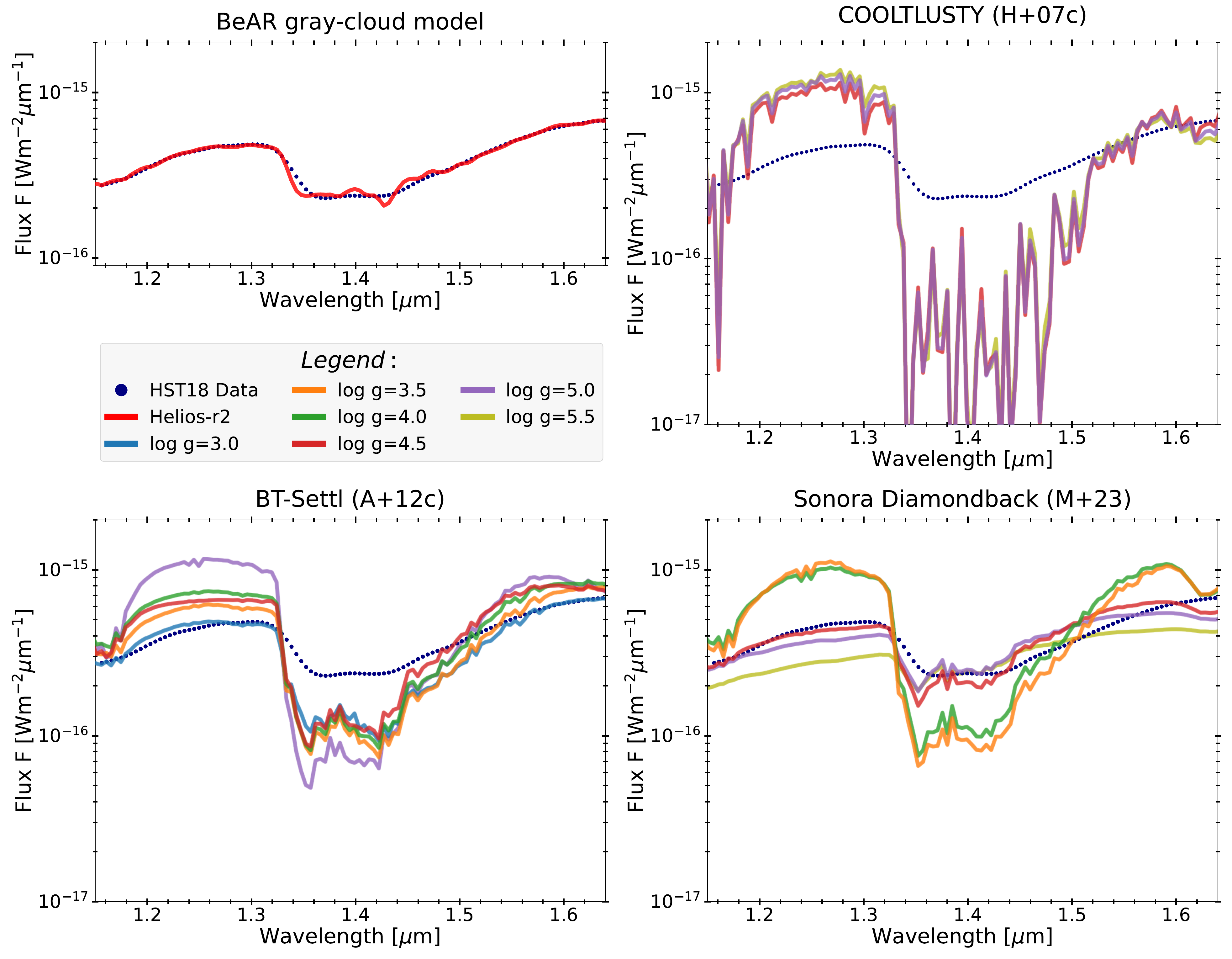}
    \caption{Collection of model spectra with an effective temperature of 1100~K and varying surface gravity values for the three legacy model grids (\texttt{COOLTLUSTY} from H+07c: \cite{Hubeny2007ApJ...669.1248H}, \texttt{BT-Settl} by A+12: \cite{Allard2012EAS....57....3A}, and \texttt{Sonora Diamondback} by M+23: \cite{Morley2024arXiv240200758M}, with a fixed [M/H] of 0.0 and $f_\mathrm{sed}=2$), as well as the best-fit spectra of the \texttt{BeAR} free-chemistry retrieval analyses with a gray-cloud model. Data is shown for the HST 2018 maximum brightness spectrum as dark blue dots with associated uncertainties.}
    \label{fig:Spectra_Comparison}
\end{figure*}

We next investigate whether performing the retrieval using pre-computed model grids will result in the same outcome. Using our three selected model grids (see Section \ref{sect:HELA}), we report the posterior distributions of the effective temperature $T_{\rm eff}$, surface gravity $\log{g}$, and radius scaling factor $f$ in Fig.~\ref{fig:ML_grids}, as well as metallicity [M/H] and cloud parameter $f_{sed}$ for \texttt{Sonora Diamondback} in Fig.~\ref{fig:ML_grids_add}. Consistent with our standard Bayesian analysis, we find that these retrieved quantities are invariant across the 231 spectra for which we performed retrievals (both epochs, 2018 and 2020). The only exception is the scaling factor $f$ for the model grid by \cite{Morley2024arXiv240200758M}, showing a reduced chi-square value of $\chi_r^2 = 2.773$ when fitting a straight line trough all 231 posteriors. However, physically, we do not expect to see an $\approx$10\% increase in the scaling factor $f$. This might well correspond to the masked effect of a proportional increase in the effective temperature $T_{\rm eff}$ (see Eq.~\ref{eq:radius-distance relation}). Previous work has shown that retrievals performed with the supervised machine learning method of the random forest tend to produce more conservative (broader) posterior distributions \citep{Fisher2020AJ....159..192F}. In our case, the great error bars are likely influenced by the large grid step sizes and sparsely sampled grids, which can increase uncertainties (see, e.g. \citealt{Lueber2023ApJ...954...22L}). While the broader posteriors from the random forest method reflect its inherent caution, the grid resolution and sampling density are additional factors contributing to the retrieved error bars.

Finally, Fig.~\ref{fig:Spectra_Comparison} illustrates the best-fit spectra in a montage, combining a collection of model spectra with an effective temperature of 1100~K and varying surface gravity values for the three model grids (\texttt{COOLTLUSTY} in \citealt{Hubeny2007ApJ...669.1248H}, \texttt{BT-Settl} in \citealt{Allard2012EAS....57....3A}, and \texttt{Sonora Diamondback} in \citealt{Morley2024arXiv240200758M}), as well as the best-fit spectra of the nested sampling (\texttt{BeAR}). For illustration, we have opted to analyze the spectrum corresponding to the maximum brightness observed in the HST data from the year 2018. Substantial discrepancies exist between the actual best-fit spectra and the data sets. Such differences are not visible in retrievals with \texttt{BeAR}. Furthermore, the individual grids differ from each other. For a more thorough analysis of this, we encourage reading  the discussion in \cite{Lueber2023ApJ...954...22L}.

Overall, the poor fits of the model grids to the data (in Fig.~\ref{fig:Spectra_Comparison}) suggest that these models are unsuitable for quantifying the temporal variation of atmospheric properties (as shown in Figs.~\ref{fig:ML_grids} and \ref{fig:ML_grids_add}) unless they are being used to simulate patchy or non-uniform atmospheres (as shown in, e.g., \citealt{Miles2023ApJ...946L...6M}).

\subsection{Constraining C/O and [M/H] ratios from medium-resolution NIR VLT/X-shooter spectra}
\label{subsect:Xshooter}

\begin{figure}
    \centering
    \includegraphics[width=0.82\columnwidth]{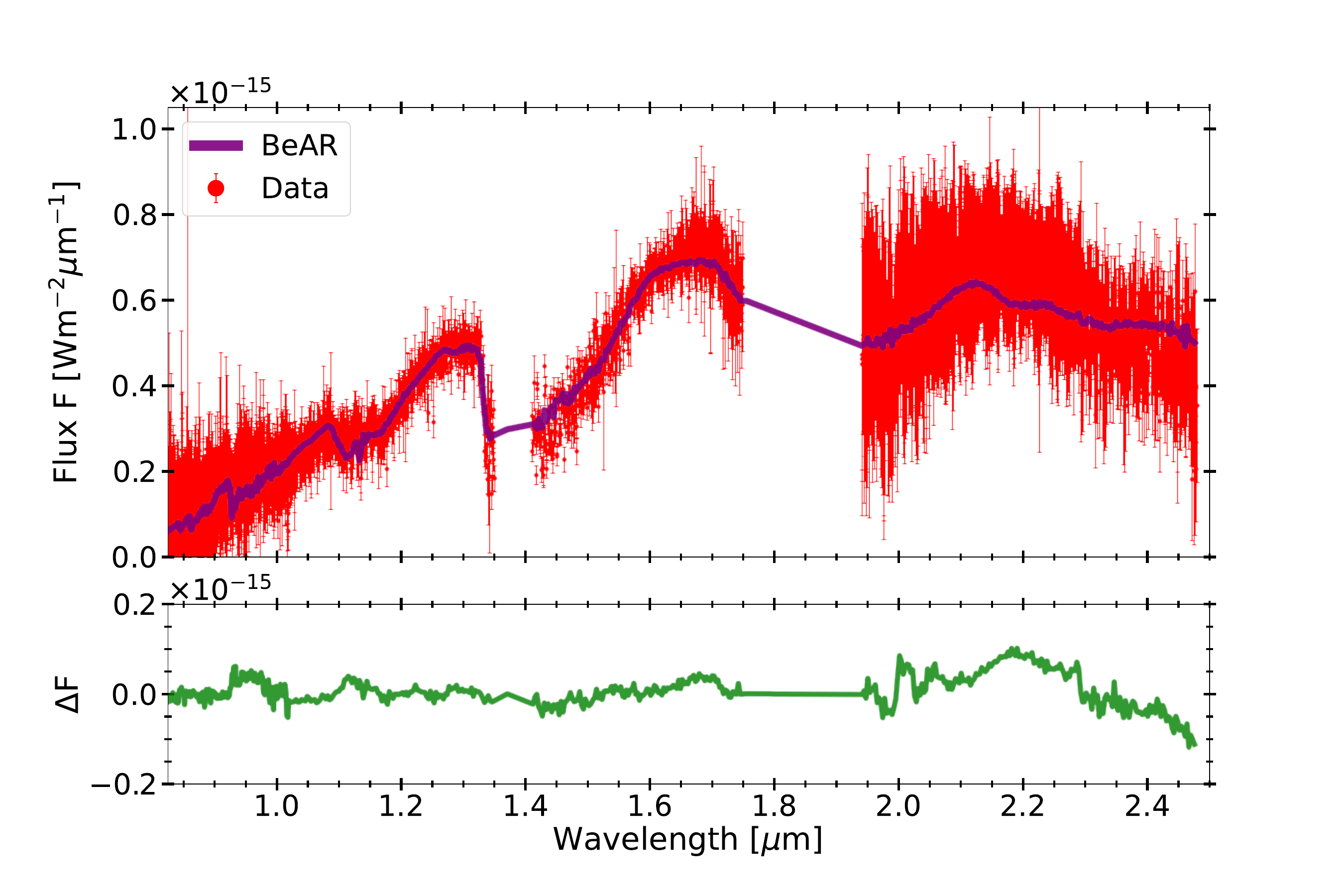}
    \caption{Posterior median spectra (F) and residuals ($\Delta$ F) associated with the free-chemistry retrieval analyses of the simultaneous $0.65-2.5 \mu$m medium-resolution VLT/X-shooter spectrum taken by \cite{Petrus2023A&A...670L...9P} with a gray-cloud model. Data are shown as red dots with associated uncertainties.}
    \label{fig:Xshooter_BF}
\end{figure}

As a complementary analysis, we considered the $0.65-2.5\,\mu$m medium-resolution ($3300 \leq$ R$_\lambda \leq 8100$) VLT/X-shooter spectrum taken by \cite{Petrus2023A&A...670L...9P}. In contrast to their study, which compared the data set to a spectral grid of cloudless \texttt{ATMO} models \citep{Tremblin2015ApJ} with the Bayesian inference tool called \texttt{ForMoSa} \citep{Petrus2023A&A...670L...9P}, we use the spectrum to calculate the C/O and [M/H] ratios from posteriors of a cloudy atmospheric retrieval with \texttt{BeAR}. Figs.~\ref{fig:Xshooter_BF} and \ref{fig:Xshooter_posteriors} show the posterior median spectra and joint posterior distributions from the free-chemistry retrieval analyses of the VLT/X-shooter spectrum with a gray-cloud model. Compared to HST retrievals of VHS~1256~b, no degeneracy between surface gravity and \ch{H2O} was found, presumably due to the significantly greater spectral resolution compared to HST/WC3, as well as the extended wavelength range, which includes gravity-sensitive lines.

The retrieved surface gravity value is slightly higher than expected with $\log{g}=4.9992_{-0.0008}^{+0.0005}$. Water abundance was constrained with a value of $\log{\ch{H2O}}=-3.37_{-0.01}^{+0.01}$, agreeing well with our previously constrained abundances of HST retrievals ($\log{\ch{H2O}}\sim10^{-3}$, when fixing the surface gravity to  $\log{g} = 4.268 \pm 0.006$ and $\log{g} = 4.45 \pm 0.03$, respectively). We note that the HST retrievals without fixed gravity values yield $\sim$1\% water abundances (Fig.~\ref{fig:VHS_retrieval_HST18max_gray} and Tab.~\ref{tab:data posteriors}). Retrieved atmospheric properties of VHS~1256~b result in a solar to super-solar C/O-ratio of C/O $= 0.77 \pm 0.49$ and a metallicity of [M/H] $= 0.53 \pm 0.27$ based on the elements contained in the molecules that are constrained in the retrieval. 

Following the enhanced data quality provided by the VLT/X-shooter spectrum, further investigation warrants a separate, in-depth study into non-uniform abundance profiles. While Bayesian model comparison strongly preferred the adoption of such non-uniform profiles of \ch{H2O} ($\ln{B_{ij}} \approx 725$), their detailed parametrisation and analysis thereof exceed the scope of the current study. Future efforts will focus on exploring the implications of these findings and leveraging the enhanced data quality provided by X-Shooter and JWST to deepen our understanding of the atmospheric composition of objects like VHS 1256b.

\subsection{HELA applied to JWST spectra}
\label{subsect:JWST}

\begin{figure}[ht]
    \centering
    \includegraphics[width=0.95\columnwidth]{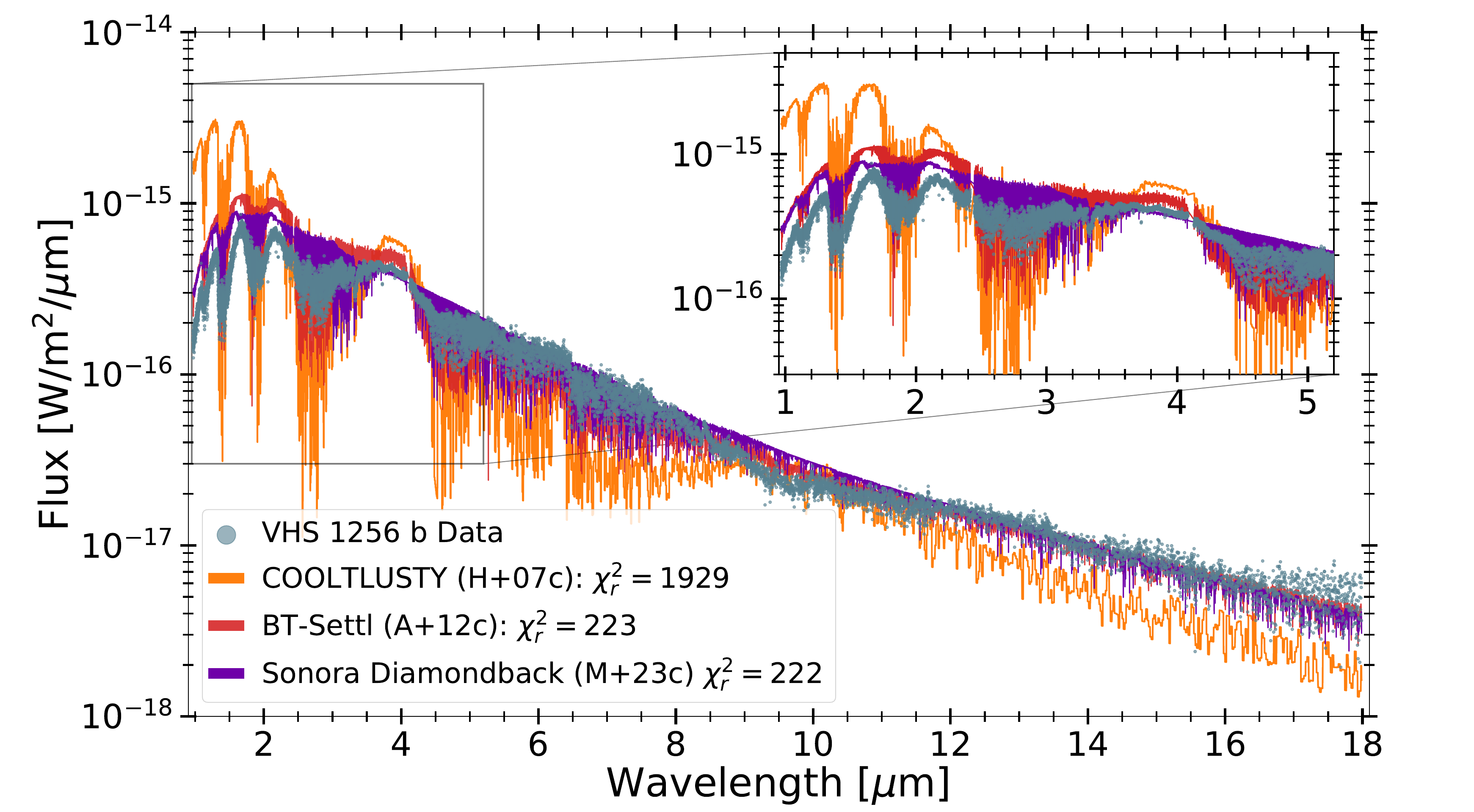}
    \vspace{0.2cm}
    \caption{JWST spectra of VHS~1256~b (gray dots), overlaid by a collection of best fit model spectra over the full wavelength coverage of the three legacy model grids (\texttt{COOLTLUSTY} from H+07c: \cite{Hubeny2007ApJ...669.1248H}, \texttt{BT-Settl} by A+12: \cite{Allard2012EAS....57....3A}, and \texttt{Sonora Diamondback} by M+23: \cite{Morley2024arXiv240200758M}.}
    \vspace{-0.2cm}
    \label{fig:HELA_JWST_BF}
\end{figure}

\begin{figure*}[ht!]
    \centering
    \includegraphics[width=0.85\textwidth]{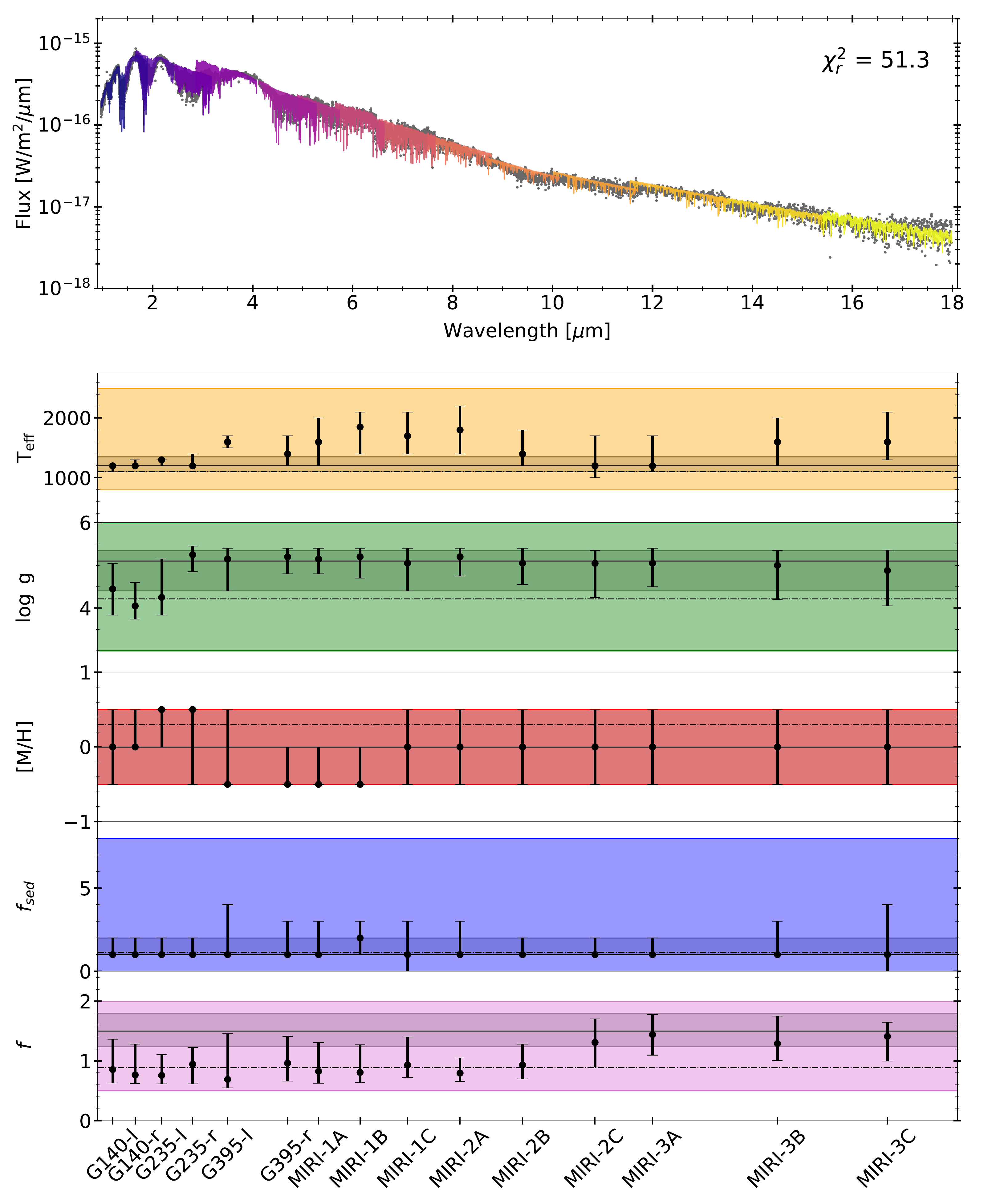}
    \caption{JWST spectra of VHS~1256~b (dark gray dots), overlaid by the \texttt{Sonora Diamondback} \citep{Morley2024arXiv240200758M} best fit model spectra for each of the 15 independent spectral windows. Corresponding posteriors for effective temperature T$_\mathrm{eff}$, surface gravity $\log{g}$, metallicity [M/H], cloud parameter $f_\mathrm{sed}$, and scaling factor $f$ are presented in the five lower panels, where the parameter range of the grid itself is indicated by the colored area. The posterior values and corresponding errors when using the entire spectrum are shown by the gray line and shaded gray area, respectively. Dashed lines represent the corresponding averaged values from the machine learning retrievals with HST data (see Figs.~\ref{fig:ML_grids} and \ref{fig:ML_grids_add}).}
    \label{fig:HELA_JWST_BF_Diamondback}
\end{figure*}

The latest publications on VHS~1256~b now focus on the JWST spectrum collected by \cite{Miles2023ApJ...946L...6M}. They observed VHS~1256~b with JWST's NIRSpec IFU and MIRI MRS modes, generating a spectrum spanning 1-18 microns at a $\sim$1000 - 3700 resolution. A forward modeling approach was used \citep{Petrus2024ApJ...966L..11P}, as well as the first atmospheric retrievals applied to the JWST data of a brown dwarf (Whiteford et al., in prep.). Both studies highlight the new challenges we are facing with JWST and indicate possible next steps towards a complete picture of VHS~1256~b. We now apply our random forest approach with HELA to analyze the JWST data.

Performing the random forest retrievals using pre-computed model grids result in best-fit spectra, shown in Fig.~\ref{fig:HELA_JWST_BF}. None of the model grids considered fit the combined JWST spectrum, as indicated by the reduced chi-square values that greatly exceed unity. Despite large deviations, the cloudy grid \texttt{Sonora Diamondback} \citep{Morley2024arXiv240200758M} provides the best fit. Following the approach in \cite{Petrus2024ApJ...966L..11P}, examining the spectra of the different instruments individually yields significantly better fits, which we show for \texttt{Sonora Diamondback} in Fig.~\ref{fig:HELA_JWST_BF_Diamondback} and \texttt{BT-Settl} in Fig.~\ref{fig:HELA_JWST_BF_Allard}. This goes along with the fact that the posteriors, effective temperature, surface gravity, and scaling factor show a wavelength dependence.

\section{Discussion}
\label{sect:Discussion}

Despite $\sim 30\%$ variations in the emergent flux of VHS~1256~b, its retrieved atmospheric properties are consistent within errors when just considering HST/WFC3 data. This implies the existence of two potential scenarios: either the properties remain invariant over time, or our observations and retrievals lack the sensitivity required to detect any temporal variability. Such substantial variations have never been definitively observed in exoplanetary atmospheres (e.g., \citealt{Agol2010ApJ...721.1861A, Apai2016ApJ...820...40A, Biller2021}). Good agreement was found between retrieved water abundances for HST and VLT/X-shooter using \texttt{BeAR} (Figs.~\ref{fig:Comparison_GN_Chi2}, \ref{fig:Comparison_GN_445_Chi2}, and \ref{fig:Xshooter_posteriors}) and retrieved parameters from the machine-learning retrievals using HELA for HST and JWST data (Figs.~\ref{fig:ML_grids}, \ref{fig:ML_grids_add}, \ref{fig:HELA_JWST_BF_Diamondback}, and \ref{fig:HELA_JWST_BF_Allard}). Additionally, our analysis of JWST spectra indicates consistency in retrieved properties across different wavelength channels, further strengthening the reliability of our findings. This bodes well for retrieval analyses that will be performed on next-generation spectra of exoplanetary atmospheres, suggesting that retrieval outcomes will be robust within the applied framework. JWST's broader wavelength coverage will likely enable more robust, time-resolved retrievals, enhancing our understanding of atmospheric variability. Heretofore, a JWST Cycle 2 proposal (Program ID: GO \#3375, PI: Whiteford) has been accepted to observe $1-14 \,\mu$m spectra of VHS 1256b in time series mode to study its variability. Our study has already found that both HST and VLT/X-shooter data necessitate vertically varying abundance profiles of water to model the spectra accurately.
Consequently, the choice of the retrieval method and its parametrisation will become even more crucial in the future. Certainly, the final word will come from performing a detailed retrieval analysis on these spectra. Hence, it will be necessary to test future retrievals for non-uniform abundance profiles akin to those exemplified in this study or to employ self-consistent models as executed by \cite{Rowland2023ApJ...947....6R} to ensure capture of the full complexity imprinted in the data. With the broader wavelength range covered by the JWST, it will become essential not only for \ch{H2O} but also for molecules that are expected to vary even more with pressure, as for example \ch{FeH}, \ch{TiO} or \ch{CH4} \citep{Rowland2023ApJ...947....6R}.

A worrying outcome of Fig.~\ref{fig:ML_grids} is that the retrieved surface gravities from machine learning grid-based retrievals are markedly different: $\log{g} \approx 5$ (\texttt{COOLTLUSTY}) versus $\log{g} \approx 3.2$ (\texttt{BT-Settl}) versus $\log{g} \approx 4.2$ (\texttt{Sonora Diamondback}). Consistent with previous work (e.g., \citealt{Oreshenko2020AJ} and \citealt{Lueber2023ApJ...954...22L}), this suggests that the retrieved surface gravity is somewhat model-dependent, unlike for the effective temperature.  For comparison, \cite{Gauza2015ApJ...804...96G} derived $\log{g}$ = 4.24 $\pm$ 0.35, and  \cite{Miles2018ApJ...869...18M} derived $\log{g}$ = 3.2 based on the use of \texttt{PHOENIX} model grids.

An even more direct comparison can be made by looking at derived values by \cite{Petrus2024ApJ...966L..11P}), where they compared parameters from fits across the full JWST wavelength range (0.97 to 18.02 $\mu$m) and from 15 distinct spectral windows, combining NIRSpec and MIRI spectra in the full range and independently fitting smaller segments in the windowed approach. Values reported by \cite{Petrus2024ApJ...966L..11P} are comparable to our study, as they found $\log{g} = 3.50 \pm 0.01$ (full range) and $3.62 \pm 0.60$ (windows) for the \texttt{BT-Settl} model, as well as $\log{g} < 3.50$ (full range) and $4.50 \pm 0.60$ (windows) for the \texttt{Sonora Diamondback} model. The \texttt{COOLTLUSTY} model was not used in their study.

Since it is difficult to decipher how each model grid is constructed thoroughly, we have not attempted to track down the source of this discrepancy, which will be the subject of future work. In general, model grids containing a large range of surface gravity values tend to result in predictions at the lower end of the prior range.  Another opportunity for future work is to run general circulation models of brown dwarfs \citep{Showman2013ApJ...776...85S} focusing on the temporal variation of their atmospheric properties.

\begin{acknowledgements}
A.L. and K.H. acknowledge partial financial support from the Swiss National Science Foundation and the European Research Council (via a Consolidator Grant to K.H.; grant number 771620), as well as administrative support from the Center for Space and Habitability (CSH). Calculations were performed on UBELIX (\url{https://www.id.unibe.ch/hpc)}, the HPC cluster at the University of Bern. J. M. V. acknowledges support from a Royal Society - Science Foundation Ireland University Research Fellowship (URF$\backslash$R1$\backslash$221932).
\end{acknowledgements}

\bibliographystyle{aa}
\bibliography{references.bib}

\onecolumn

\begin{appendix}
\section{Additional Posterior Tables and Figures}
\label{sect: Appendix}
\begin{figure*}[ht!]
    \centering
    \includegraphics[width=\textwidth]{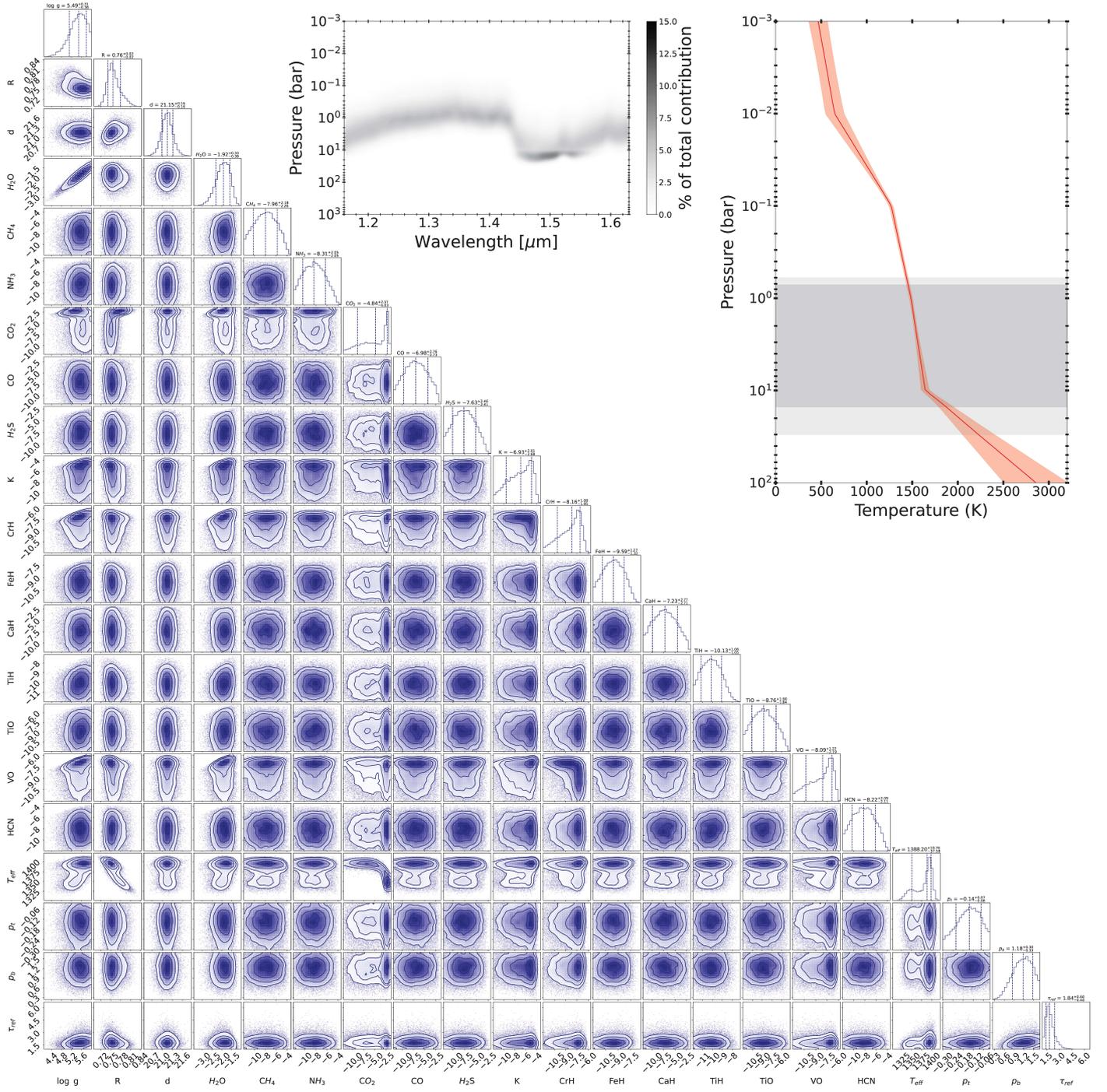}
    \caption{Joint posterior distributions from free-chemistry retrieval analyses of the HST 2018 maximum brightness spectrum with a gray-cloud model. Note that the surface gravity is retrieved for and not a fixed value. The vertical dashed lines correspond to the median parameter values and their 1$\sigma$ uncertainties. Accompanying each montage of joint posterior distributions is the retrieved median temperature–pressure profile and its associated 1$\sigma$ uncertainties. The gray horizontal bar indicated the retrieved cloud pressures (from $p_t$ and $p_b$). The effective temperature $T_\mathrm{eff}$ is the only parameter that is determined via post-processing. Additionally, the contribution function is displayed in the top center panel, showing each pressure’s contribution percentage to the flux at each wavelength as black shading.}
    \label{fig:VHS_retrieval_HST18max_gray}
\end{figure*}

\begin{figure*}[h!]
    \centering
    \includegraphics[width=0.49\columnwidth]{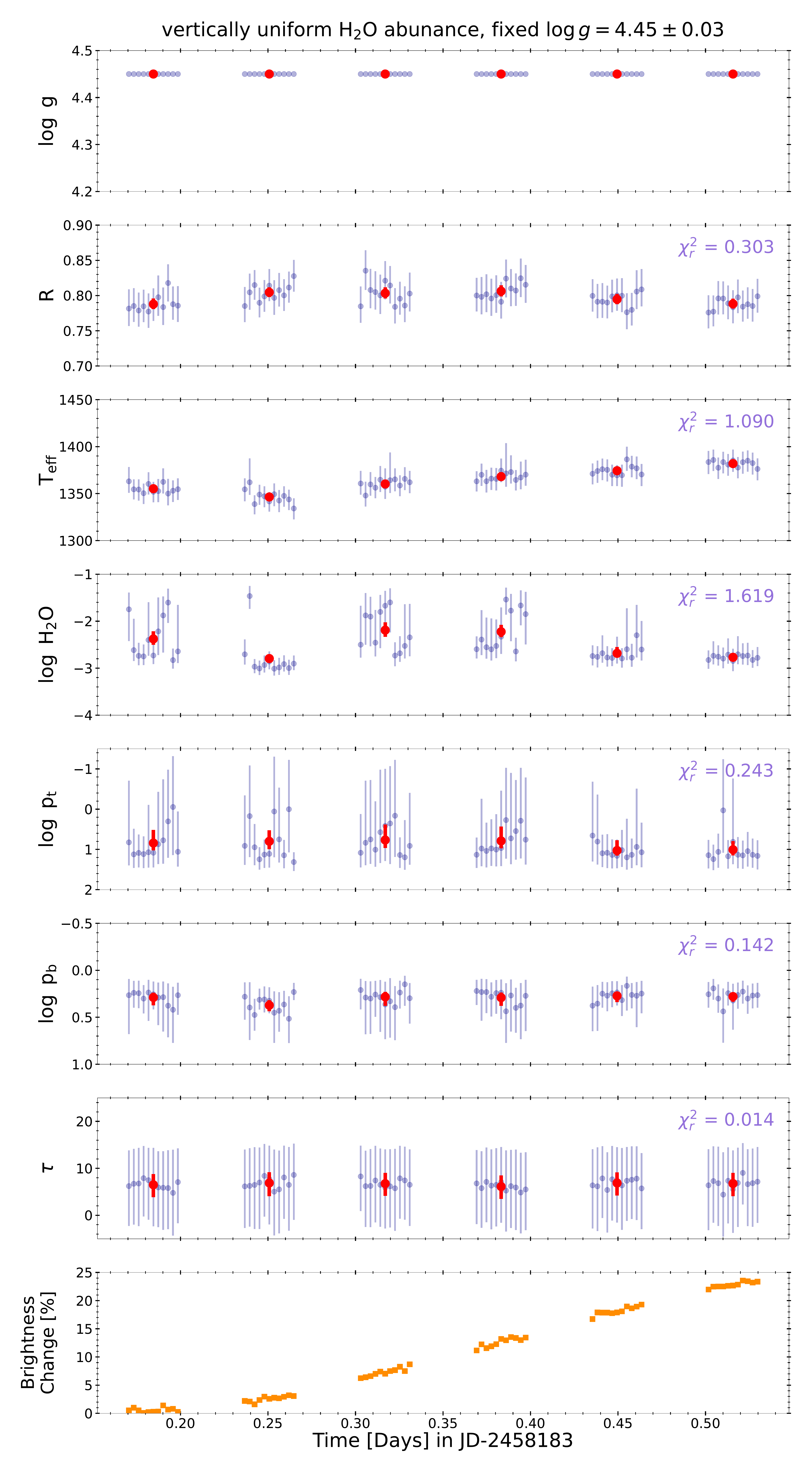}
    \includegraphics[width=0.49\columnwidth]{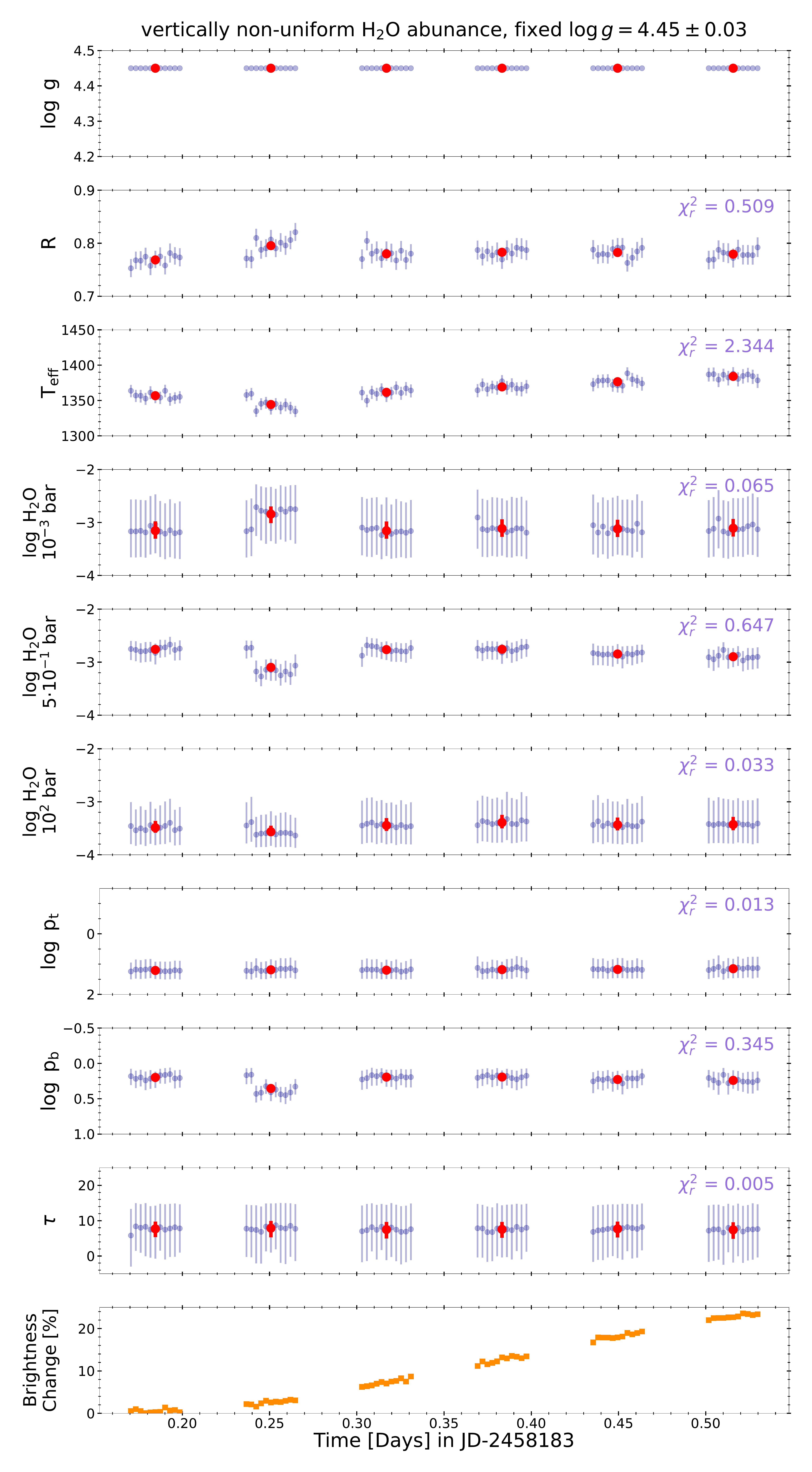}
    \caption{HST18 \texttt{BeAR} posteriors with a fixed surface gravity of $\log{g}=4.45 \pm 0.03$ and vertically uniform (left column) vs. non-uniform (right column) \ch{H2O} abundances as a sequence over all 6 orbits in 2018. Each orbit consist of 11 intra-orbit spectra. Orbit-averaged posteriors are shown in red. The reduced chi-square values correspond to a fit of a straight line through all posteriors of a given parameter. The lowest panel represents the brightness change in G141 broad band, normalized by $3.0 \times 10^{-13}$~erg/s/cm$^2$/$\mu$m (data taken from \citealt{Bowler2020ApJ...893L..30B} and \citealt{Zhou2022AJ....164..239Z}).}
    \label{fig:Comparison_GN_445_Chi2}
\end{figure*}

\begin{figure*}
    \centering
    \includegraphics[width=\textwidth]{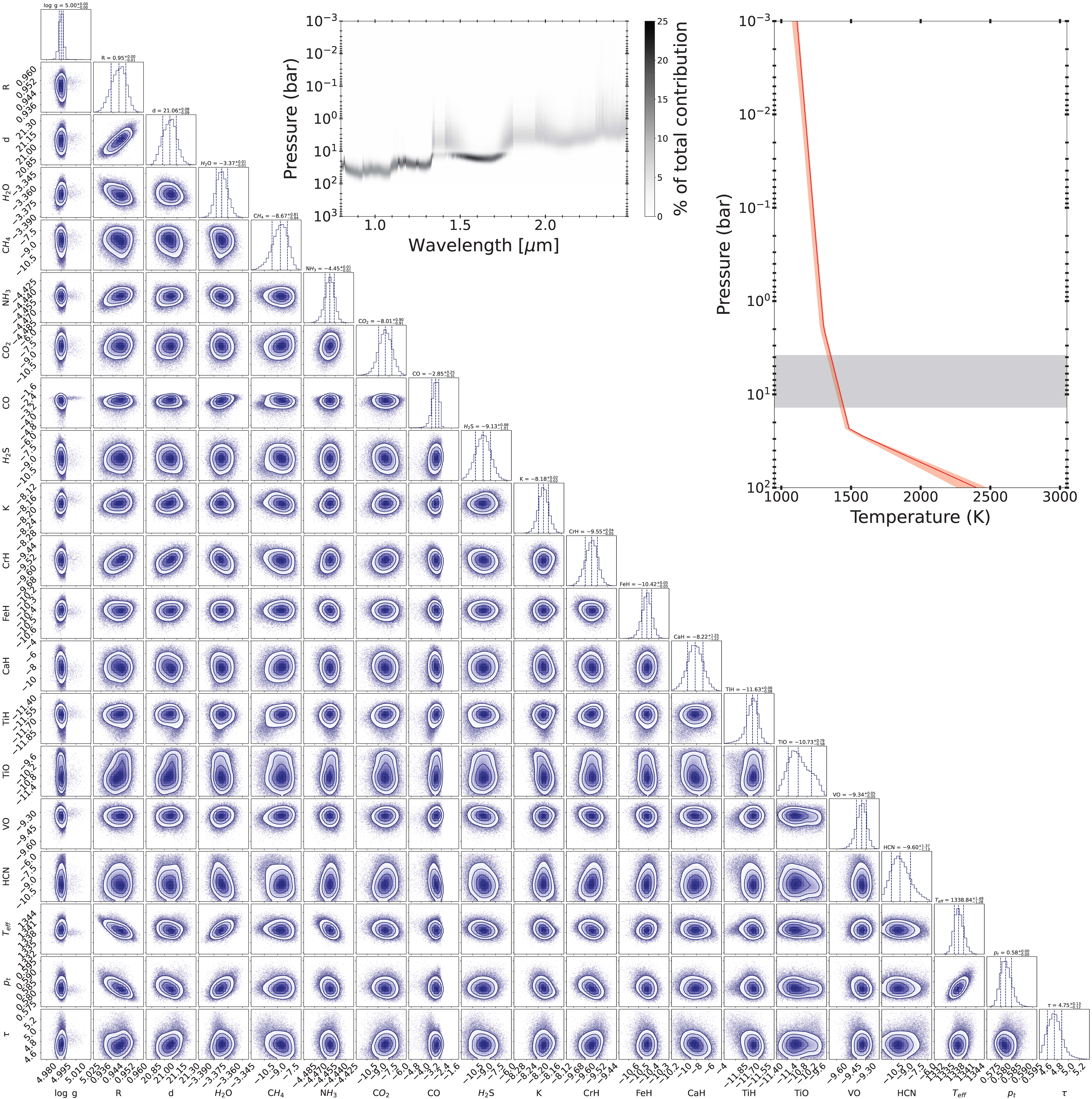}
    \caption{Joint posterior distributions from free-chemistry retrieval analyses of the VLT/X-shooter spectrum taken by \cite{Petrus2023A&A...670L...9P} with a gray-cloud model. The vertical dashed lines correspond to the median parameter values and their 1 $\sigma$ uncertainties. Accompanying each montage of joint posterior distributions is the retrieved median temperature–pressure profile and its associated 1 $\sigma$ uncertainties. The gray horizontal bar indicated the retrieved cloud pressures (from $p_t$ and $p_b$). The effective temperature $T_\mathrm{eff}$ is the only parameter that is determined via post-processing. Additionally, the contribution function is displayed in the top center panel, showing each pressure’s contribution percentage to the flux at each wavelength as black shading.}
    \label{fig:Xshooter_posteriors}
\end{figure*}

\begin{figure*}[h!]
    \centering
    \includegraphics[width=0.85\textwidth]{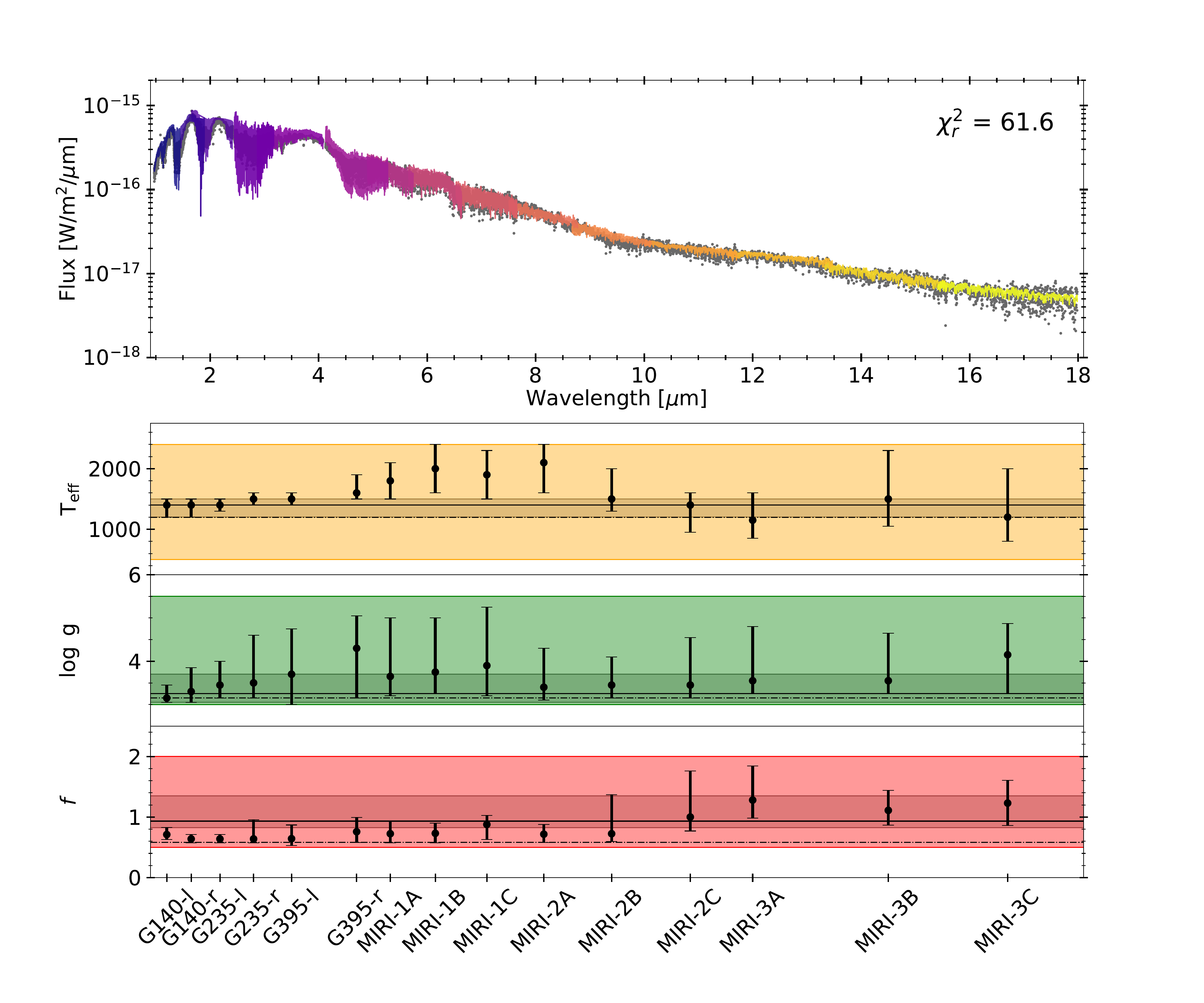}
    \caption{JWST spectra of VHS~1256~b (dark gray dots), overlaid by the \texttt{BT-Settl} \citep{Allard2011ASPC..448...91A} best fit model spectra for each of the 15 independent spectral windows. Corresponding posteriors for effective temperature T$_\mathrm{eff}$, surface gravity $\log{g}$, and scaling factor $f$ are presented in the five lower panels, where the parameter range of the grid itself is indicated by the colored area. The posterior values and corresponding errors when using the entire spectrum are shown by the gray line and shaded gray area, respectively. Dashed lines represent the corresponding averaged values from the machine learning retrievals with HST data (see Fig.~\ref{fig:ML_grids} and \ref{fig:ML_grids_add}).}
    \label{fig:HELA_JWST_BF_Allard}
\end{figure*}

For completeness, Tabs.~\ref{tab:data posteriors} and \ref{tab:Bayes data} record the outcomes of the standard Bayesian retrieval analysis using \texttt{BeAR} for the orbit-averaged HST spectra from 2018 and 2020, and the corresponding summary of Bayesian statistics, respectively. Fig.~\ref{fig:VHS_retrieval_HST18max_gray} represents the joint posterior distributions from free-chemistry retrieval analyses of the HST 2018 maximum brightness spectrum with a gray-cloud model. Fig.~\ref{fig:Comparison_GN_445_Chi2} shows the retrieved atmospheric properties across the 66 different HST spectra from 2018, which are constructed from 6 different orbits of VHS~1256~b. This Figure is analogous to Fig.~\ref{fig:Comparison_GN_Chi2}, but fixing the surface gravity value of $\log{g} = 4.45 \pm 0.03$. In Fig.~\ref{fig:Xshooter_posteriors}, the joint posterior distributions from the free-chemistry retrieval analyses of the VLT/X-shooter spectrum taken by \cite{Petrus2023A&A...670L...9P} and using a gray-cloud model is shown. Fig.~\ref{fig:HELA_JWST_BF_Allard} then represents the best-fit spectra and posteriors from performing random forest retrievals applied to the JWST spectra of VHS 1256 b, looking at each instrument individually and using the pre-computed model grid \texttt{BT-Settl} \citep{Allard2011ASPC..448...91A}.

\begin{table*}[h!]
\centering
\caption{Summary of orbit-averaged retrieval outcomes for the brown dwarf VHS~1256~b.}
\label{tab:data posteriors}
\resizebox{0.9\textwidth}{!}{
\tiny
\begin{tabular}{rrrrrrrr}
\hline
Model  & Parameter & HST18 ave & HST18 max & HST18 min & HST20 ave & HST20 max & HST20 min \\
\hline
cloud-free & $\log g$ [cm/s$^2$] & $5.37_{-0.32}^{+0.21}$& $5.19_{-0.37}^{+0.23}$ & $5.04_{-0.32}^{+0.22}$ & $5.02_{-0.75}^{+0.43}$& $5.19_{-0.39}^{+0.26}$& $4.45_{-0.88}^{+0.70}$\\
gray & $\log g$ [cm/s$^2$] &$5.67_{-0.31}^{+0.22}$	& $5.65_{-0.28}^{+0.21}$ & $5.64_{-0.29}^{+0.21}$ & $5.41_{-0.55}^{+0.33}$& $4.57_{-0.73}^{+0.82}$& $5.46_{-0.36}^{+0.27}$\\
non-gray & $\log g$ [cm/s$^2$] & $5.73_{-0.23}^{+0.17}$& $5.45_{-0.28}^{+0.21}$ & $5.33_{-0.35}^{+0.34}$& $5.36_{-0.61}^{+0.34}$ & $5.23_{-0.40}^{+0.28}$& $5.30_{-1.36}^{+0.23}$\\
\hline
cloud-free & $R$ [R$_\mathrm{J}$] & $0.76_{-0.01}^{+0.01}$ & $0.76_{-0.01}^{+0.01}$ & $0.76_{-0.01}^{+0.01}$& $0.78_{-0.01}^{+0.01}$ & $0.78_{-0.01}^{+0.01}$ & $0.79_{-0.01}^{+0.01}$ \\
gray & $R$ [R$_\mathrm{J}$] & $0.75_{-0.01}^{+0.01}$ & $0.75_{-0.01}^{+0.01}$ & $0.75_{-0.01}^{+0.01}$& $0.77_{-0.01}^{+0.01}$ & $0.78_{-0.01}^{+0.01}$ & $0.77_{-0.01}^{+0.01}$ \\
non-gray & $R$ [R$_\mathrm{J}$] & $0.75_{-0.01}^{+0.01}$ & $0.75_{-0.01}^{+0.01}$ & $0.75_{-0.01}^{+0.01}$& $0.77_{-0.01}^{+0.01}$ & $0.77_{-0.01}^{+0.01}$ & $0.76_{-0.01}^{+0.02}$  \\
\hline
cloud-free & $d$ [pc] & $21.17_{-0.20}^{+0.18}$ & $21.15_{-0.18}^{+0.17}$ &	$21.16_{-0.18}^{+0.17}$& $21.14_{-0.18}^{+0.18}$& $21.15_{-0.18}^{+0.17}$ & $21.16_{-0.17}^{+0.16}$ \\
gray & $d$ [pc] & $21.15_{-0.18}^{+0.18}$ & $21.14_{-0.16}^{+0.16}$ &	$21.13_{-0.16}^{+0.16}$& $21.14_{-0.18}^{+0.18}$& $21.14_{-0.17}^{+0.17}$ & $21.15_{-0.16}^{+0.16}$ \\
non-gray & $d$ [pc] & $21.14_{-0.17}^{+0.17}$ & $21.15_{-0.17}^{+0.17}$ & $21.15_{-0.16}^{+0.16}$& $21.14_{-0.18}^{+0.17}$ & $21.14_{-0.16}^{+0.16}$& $21.13_{-0.15}^{+0.15}$ \\
\hline
cloud-free & $\log$ \ch{H2O} & $-1.35_{-0.36}^{+0.23}$	& $-1.46_{-0.41}^{+0.29}$ & $-1.39_{-0.34}^{+0.24}$& $-1.85_{-0.84}^{+0.53}$ & $-1.59_{-0.43}^{+0.32}$ & $-2.34_{-0.93}^{+0.80}$ \\
gray & $\log$ \ch{H2O} & $-1.74_{-0.35}^{+0.28}$	& $-1.61_{-0.30}^{+0.26}$ & $-1.61_{-0.32}^{+0.26}$ & $-1.69_{-0.60}^{+0.39}$ & $-2.53_{-0.75}^{+0.90}$ & $-1.69_{-0.38}^{+0.32}$ \\
non-gray & $\log$ \ch{H2O} & $-1.76_{-0.25}^{+0.24}$ & $-1.40_{-0.30}^{+0.24}$ & $-1.39_{-0.31}^{+0.23}$& $-1.70_{-0.65}^{+0.41}$ & $-1.71_{-0.44}^{+0.34}$ & $-1.71_{-1.28}^{+0.27}$\\
\hline
cloud-free & $T_\mathrm{{eff}}$ [K] & $1368.60_{-7.65}^{+6.02}$	& $1382.44_{-9.09}^{+6.98}$ & $1360.17_{-7.24}^{+6.21}$ & $1381.93_{-11.22}^{+7.17}$& $1387.37_{-8.87}^{+5.73}$& $1370.52_{-10.14}^{+6.50}$\\
gray & $T_\mathrm{{eff}}$ [K] & $1377.48_{-9.52}^{+6.59}$	& $1388.52_{-9.98}^{+6.84}$ & $1365.97_{-7.46}^{+5.71}$& $1385.26_{-13.31}^{+6.76}$& $1380.35_{-18.38}^{+10.52}$& $1378.29_{-7.43}^{+4.83}$\\
non-gray & $T_\mathrm{{eff}}$ [K] & $1374.62_{-8.58}^{+6.97}$ & $1381.85_{-8.04}^{+6.35}$ & $1358.98_{-6.37}^{+5.78}$& $1386.35_{-12.10}^{+6.96}$ & $1388.52_{-10.05}^{+5.84}$ & $1379.32_{-11.67}^{+6.43}$  \\
\hline
cloud-free & $\log p_{\mathrm{t}}$ [bar] & - & - & - & - & - & - \\
gray & $\log p_{\mathrm{t}}$ [bar] & $-0.17_{-0.08}^{+0.08}$	& $-0.14_{-0.08}^{+0.07}$ & $-0.18_{-0.07}^{+0.07}$& $-0.16_{-0.08}^{+0.08}$& $-0.16_{-0.08}^{+0.08}$ & $-0.15_{-0.08}^{+0.07}$ \\
non-gray & $\log p_{\mathrm{t}}$ [bar] & $-0.18_{-0.07}^{+0.08}$	& $-0.18_{-0.07}^{+0.08}$ & $-0.21_{-0.06}^{+0.07}$& $-0.18_{-0.08}^{+0.08}$ & $-0.18_{-0.07}^{+0.08}$ & $-0.18_{-0.07}^{+0.07}$ \\
\hline
cloud-free & $\log p_{\mathrm{b}}$ [bar] & - &- & - & - & - & - \\
gray & $\log p_{\mathrm{b}}$ [bar] & $1.27_{-0.31}^{+0.26}$ & $1.30_{-0.27}^{+0.23}$ & $1.31_{-0.25}^{+0.23}$& $0.96_{-0.45}^{+0.45}$& $0.82_{-0.33}^{+0.49}$ & $1.26_{-0.33}^{+0.26}$ \\
non-gray & $\log p_{\mathrm{b}}$ [bar] & $1.36_{-0.28}^{+0.22}$ & $1.27_{-0.41}^{+0.27}$ & $1.15_{-0.95}^{+0.36}$& $1.15_{-0.41}^{+0.34}$& $1.20_{-0.37}^{+0.31}$ & $1.25_{-0.38}^{+0.28}$\\
\hline
cloud-free & $\tau$ & - & - & - & - & - & - \\
gray & $\tau$ & $1.94_{-0.42}^{+0.47}$ & $1.73_{-0.31}^{+0.29}$ & $1.53_{-0.26}^{+0.32}$& $1.99_{-0.74}^{+0.65}$ & $2.00_{-0.69}^{+0.82}$ & $1.81_{-0.35}^{+0.37}$ \\
non-gray & $\tau$ & $3.23_{-1.56}^{+2.98}$ & $3.14_{-1.50}^{+2.95}$ & $3.22_{-1.54}^{+2.86}$& $3.17_{-1.57}^{+3.18}$ & $3.25_{-1.58}^{+2.90}$ & $3.15_{-1.46}^{+2.67}$ \\
\hline
cloud-free & $\log Q_0$  & - & - & - & -& -& - \\
gray & $\log Q_0$ & - &- & - & -& -& - \\
non-gray & $\log Q_0$ &  $6.96_{-4.79}^{+6.83}$ & $6.00_{-4.17}^{+5.63}$ & $6.38_{-4.43}^{+6.45}$ & $6.57_{-4.55}^{+6.93}$& $6.13_{-4.24}^{+6.48}$ &$5.43_{-3.65}^{+4.67}$ \\
\hline
cloud-free & $a_0$ & - & - &- & - & - & - \\
gray & $a_0$ & - & - &-  & - & - & - \\
non-gray & $a_0$ & $9.55_{-6.91}^{+25.65}$ & $10.21_{-7.48}^{+27.30}$ & $10.74_{-7.84}^{+27.94}$& $9.95_{-7.40}^{+29.11}$ & $9.61_{-6.88}^{+26.36}$ & $10.90_{-7.77}^{+26.08}$  \\
\hline
cloud-free & $\log a$ [$\mu$m] & - & - & - & - & - & - \\
gray  & $\log a$ [$\mu$m] & - & - & - &-  & - & - \\
non-gray & $\log a$ [$\mu$m] & $0.51_{-0.02}^{+0.02}$ & $0.52_{-0.03}^{+0.06}$ & $0.54_{-0.04}^{+0.14}$ & $0.59_{-0.06}^{+0.07}$ & $0.56_{-0.05}^{+0.06}$ & $0.52_{-0.02}^{+0.04}$ \\
\hline
\end{tabular}}
\end{table*}

\begin{table*}[hb!]
\centering
\caption{Summary of Bayesian statistics for the brown dwarf VHS~1256~b orbit-averaged retrievals.}
\label{tab:Bayes data}
\resizebox{0.9\textwidth}{!}{
\tiny
\begin{tabular}{cccccccc}
\hline
Model  & Parameter & HST18 ave & HST18 max & HST18 min & HST20 ave & HST20 max & HST20 min \\
\hline
cloud-free & evidence & 4203.28 & 4180.01 & 4205.45 & 4222.25 & 4206.07 & 4225.31 \\
gray & evidence & 4213.76 & 4187.01 & 4217.81 & 4225.34 & 4209.46 & 4230.59 \\
non-gray & evidence & 4207.17 & 4180.92 & 4207.09 & 4223.49 & 4206.19 & 4225.51 \\
\hline
\hline
gray vs. cloud-free & $\ln B_{ij}$ & 10.48 & 7.09 & 12.36 & 3.09 & 3.39 & 5.28 \\
non-gray vs. cloud-free & $\ln B_{ij}$ & 3.89 & 0.90 & 1.64 & 1.24 & 0.12 & 0.20 \\
gray vs. non-gray & $\ln B_{ij}$ & 6.60 & 6.19 & 10.72 & 1.85 & 3.27 & 5.08 \\
\hline
\end{tabular}}
\end{table*}

\end{appendix}

\end{document}